\documentclass[aps,prc,twocolumn,showpacs,amsmath,amssymb,nofootinbib]{revtex4}
\usepackage {lscape}
\usepackage{graphicx}
\usepackage{amsmath}
\usepackage{rotating}

\usepackage[bookmarks=true]{hyperref}
\usepackage{color}
\usepackage{longtable}

  \hypersetup{
   colorlinks=true,
   pdfborder={0 0 0},
   citecolor=blue,
   linkcolor=blue,
   urlcolor=blue
 }

\begin{document}

\title{First-principles results for electromagnetic properties of $sd$ shell nuclei}

\author{ Archana Saxena\footnote{archanasaxena777@gmail.com} and Praveen C. Srivastava\footnote{pcsrifph@iitr.ac.in} }
\address{Department of Physics, Indian Institute of Technology Roorkee, Roorkee
247 667, India}

\date{\hfill \today}

\begin{abstract}

In this work we present $ab~initio$ shell-model calculations for electric quadrupole moments and magnetic dipole moments of $sd$ shell
nuclei using valence-space Hamiltonians derived with two $ab~initio$ approaches: the in-medium similarity renormalization group (IM-SRG)
and the coupled-cluster effective interaction (CCEI).  Results are in a reasonable
agreement with the available experimental data as well as with the results from the phenomenological USDB effective interaction. This work will add
more information to the available $ab~initio$ results for the spectroscopy of $sd$ shell nuclei.

\end{abstract}

\maketitle
\section{Introduction}

The study of atomic nuclei using first principles is an important topic in the nuclear structure physics.
The anomalous behavior for nuclei close to the drip line can now be explained using $ab~initio$ approaches.
The inclusion of three-body forces was found to be crucial for explaining the exact location of the drip-line for
oxygen and calcium isotopes \cite{otsu1,otsu2}. Although $ab~initio$ calculations are difficult for heavier nuclei,
recently spectroscopy of $sd$ shell nuclei using different $ab~initio$ approaches has been reported in the literature.
Using the in-medium similarity renormalization group (IM-SRG) approach, $ab~initio$ predictions for the ground and excited states of
doubly open-shell $sd$ nuclei have been reported in Ref. \cite{stroberg}. Also $ab~initio$ coupled-cluster effective
interaction (CCEI) was derived and used to calculate the levels in $p$ and $sd$ shell nuclei successfully \cite{jan1,jan2,sri}.
A mass-dependent effective Hamiltonian in a $0\hbar\Omega$ model space for the $sd$ shell nuclei, starting from a no core shell
model Hamiltonian in a $4\hbar\Omega$ model space with 
the realistic J-matrix Inverse Scattering Potential, fitted
to nuclei with masses up to A = 16 (JISP16) and chiral 
next-to-next-to-next-to leading order (N3LO) N N interactions, has been reported
in Ref. \cite{Dikmen}. The recent experimental results of $^{24}$F have been theoretically interpreted with IM-SRG in Ref. \cite{24F_imsrg}
and a coupled-cluster interpretation has been presented for recently populated levels in $^{25}$F \cite{25F_cc}.  In all these papers,
the focus was on the spectroscopy of $p$ and $sd$ shell nuclei.

In the present work, our motivation is to test the $ab~initio$ Hamiltonians derived from the two approaches, IM-SRG and CCEI, by calculating
electromagnetic properties of $sd$ shell nuclei.  The results of this work will add to the earlier studies in
Refs. \cite{stroberg,jan1,jan2,Dikmen}, where only spectroscopic properties (spins, parities, and level energies)
of these nuclei were reported. We compare our results to the available experimental data as well as with the calculations
using the phenomenological  USDB shell-model interaction \cite{usdb}.

The paper is organized as follows. In Sec. II, we present the details about the Hamiltonians from $ab ~ initio$ approaches.
In Sec. III, we present the theoretical results along with the experimental data wherever these are available. 
Finally, a summary and conclusions are presented in Sec. IV.

\section{\bf{Hamiltonian}}

In this work, we have performed shell-model calculations for which the valence-space Hamiltonian was derived
using two modern $ab ~ initio$ approaches: in-medium similarity renormalization group (IM-SRG) \cite{stroberg} and
coupled-cluster effective interaction (CCEI) \cite{jan1}. We have also compared the results with calculations
using the phenomenological  USDB effective interaction \cite{usdb}. For the diagonalization of the matrices we have
used the shell-model code NUSHELLX \cite{Nushellx}.

Using in-medium similarity renormalization group (IM-SRG) \cite{bogner} based on chiral two- and three-nucleon interactions,
Stroberg $\it {et ~ al.}$ derived mass-dependent Hamiltonians for $sd$ shell nuclei \cite{stroberg}. 

The starting Hamiltonian $H$ is normal-ordered with respect to $|\Phi_0\rangle$  
\begin{multline}
H = E_0 +  \sum_{ij}f_{ij}\{a_i^{\dagger}a_j\}
+ \frac{1}{2!^2}\sum_{ijkl}\Gamma_{ijkl}\{a^{\dagger}_ia^{\dagger}_ja_la_k\}\\
+ \frac{1}{3!^2}\sum_{ijklmn}W_{ijklmn} \{a^{\dagger}_ia^{\dagger}_ja^{\dagger}_ka_na_ma_l\},
\end{multline}

where the normal-ordered strings of creation and annihilation operators obey 
 $\langle\Phi_0|\{a_i^{\dagger}\ldots a_j\}|\Phi_0\rangle\!=\!0$. 
 The normal-ordered zero-, one-, two- and three-body terms are $E_0$, $f_{ij}$, $\Gamma_{ijkl}$, and $W_{ijklmn}$, respectively (see
 Ref. \cite{tsukiyama} for further details).
 Here, a continuous unitary transformation, parametrized
by the flow parameter $s$, is applied to the initial
normal-ordered $A$-body Hamiltonian:

\begin{equation}
  H(s)=U^{\dag}(s)H(0)U(s)\,=H^{d}(s)+H^{od}(s),
\end{equation}

where, $H^{d}(s)$ is the diagonal part of the Hamiltonian and $H^{od}(s)$ is
the off-diagonal part of the Hamiltonian. The evolution of $H(s)$ with $s$ is given by

\begin{equation}
\frac{dH(s)}{ds}=[\eta(s),H(s)],
\end{equation}

where $\eta(s)$ is the anti-Hermitian generator of unitary transformation
\begin{equation}
\eta(s)=\frac{dU(s)}{ds}U^{\dag}(s)=-\eta^{\dag}(s).
\end{equation}

The off-diagonal matrix elements become zero as $s\rightarrow\infty$ for an appropriate value of $\eta(s)$.
Here, the $sd$ valence space is decoupled from the core and higher shells as $s\rightarrow\infty$. Now we use this resulting Hamiltonian
in the shell-model calculations with $\hbar\Omega$=24 MeV. Further details about the parameters are given in Ref. \cite{stroberg}.

There is another $ab~initio$ approach, named the coupled cluster effective interaction, which we use in this paper
for calculating the electromagnetic properties of nuclei in the $sd$ shell. 
The effective interaction developed from this approach has already
been successfully used to calculate the spectroscopy for open $sd$ shell deformed nuclei \cite{jan1}.
For the effective interaction from this approach, we have also used a Hamiltonian which is $A$-dependent:

\begin{equation}
  \label{intham}
  \hat{H_A} = \sum_{i<j}\left({({\bf p}_i-{\bf p}_j)^2\over 2mA} + \hat{V}
    _{NN}^{(i,j)}\right) + \sum_{ i<j<k}\hat{V}_{\rm 3N}^{(i,j,k)}.
\end{equation}

The $NN$ and $NNN$ parts are taken from a next-to-next-to leading order (N2LO) chiral nucleon-nucleon interaction
and a next-to-next-to-next-to leading order (N3LO) chiral three-body interaction, respectively. A cutoff scale $\Lambda=500$ MeV is used
for the $NN$ part and $\Lambda=400$ MeV for the $NNN$ part.
These interactions are constructed using the similarity transformation group method (see Ref. \cite{jan2} for further details).
In this approach, a Hartree-Fock ground state in 13 oscillator major shells with $\hbar\Omega$=20 MeV is used. The CCEI
Hamiltonian for shell-model calculations can then be expanded as 

\begin{equation}
{H_{CCEI}}=H_0^{A_c}+H_1^{A_c+1}+H_1^{A_c+2}+\cdots.
\end{equation}

Here $H_0^{A_c}$, $H_0^{A_c+1}$, and $H_0^{A_c+2}$ are called core, one-body, and two-body cluster Hamiltonians, respectively.
This expansion is known as the valence cluster expansion. Similar to this Hamiltonian,
any operator can be expanded in the valence space for the shell model calculations.

The two-body term is computed using the Okubo-Lee-Suzuki (OLS) similarity transformation. We get a non-Hermitian
effective Hamiltonian in this procedure.
The metric operator $S^{\dag}$$S$ is used to make the Hamiltonian Hermitian. The resultant effective Hermitian
Hamiltonian used for the shell model is $[S^{\dag}S]^{1/2}H_{CCEI}[S^{\dag}S]^{-1/2}$.  Here, $S$ is a matrix that diagonalizes $H_{CCEI}$.

Finally, we have also compared our $ab ~ initio$ results with the shell model calculations using the phenomenological  USDB interaction.
The USDB interaction is fitted to two-body matrix elements \cite{usdb}, originally derived from a $G$-matrix approach.
This interaction is fitted by varying 56 linear combinations of two-body matrix elements.

The shell model code NUSHELLX@MSU is a set of wrapper codes written by Brown.
It uses a proton-neutron basis.  With this code, it is possible to diagonalize $J$-scheme matrix dimensions up to $\sim$ 100 million.

\section{Results and discussions}

The magnetic dipole moment is defined as the expectation value of the dipole operator in the state
with maximum $M$ projection as

\begin{equation}
 \mu =\langle J;M=J\mid \sum_{i} g_l(i)l_{z,i} +\sum_{i} g_s(i)s_{z,i} \mid J;M =J\rangle.
\end{equation}

Here, $g_{l}$ and $g_{s}$ are the orbital and spin gyromagnetic ratios, respectively. By applying the Wigner-Eckart theorem,

\begin{equation*}
 \mu = \frac {J}{[J(J+1)(2J+1)]^{1/2}} 
\end{equation*}

\begin{equation}
 \times \langle J \mid \mid \sum_{i}  {g_l(i){\bf j_{i}}+[g_s(i)-g_{l}(i)]} {\bf {s_{i}}}  \mid \mid J \rangle.
 \end{equation}

The electric quadrupole moment operator is defined as
\begin{equation}
 Q_z= \sum_{i=1}^A {Q_Z(i)} = \sum_{i=1}^A e_i(3z_i^2-r_i^2).
\end{equation}
The spectroscopic quadrupole moment ($Q_{s}$) is defined as

\begin{equation*}
  Q_s({J})=\langle  {J},m=  {J} \mid  {Q_2^0} \mid {J},m  = {J}\rangle   \\
  \end{equation*}

\begin{equation}
=\sqrt\frac{ {J}(2 {J}-1)}{(2{J}+1)(2{J}+3)({J}+1)}\langle{J}\mid\mid  {Q} \mid\mid {J}\rangle.
  \end{equation}

We have used the harmonic-oscillator parameter $\hbar \Omega =45 A^{-1/3} -25 A^{-2/3}$ MeV for all the three calculations.
The calculated values of the electromagnetic properties of $sd$ shell nuclei
with $e_p=1.5e$, $e_n=0.5e$, and $g_s^{eff}$ =  $g_s^{free}$ using the two $ab~initio$ interactions
as well as the phenomenological  USDB shell-model interaction in the $sd$ model space, along with the experimental
data, are shown in the Tables  \ref{mag} and \ref{qad}.
In Ref. \cite{PRC78_064302}, the suitable values of $g$
factors and effective charges for $sd$ shell nuclei are given.
However, in the present work, we have compared magnetic
and quadrupole moments with two $ab~initio$ effective interactions along with phenomenological USDB effective
interaction using free-nucleon $g$ factors and standard
values of effective charges in our calculations.
The magnetic moments have been taken from Ref. \cite{stone2005} and more recent data
were obtained from a compilation maintained by Mertzimekis under the IAEA auspices \cite{IAEA https://www-nds.iaea.org/nuclearmoments/}.
Recently, the experimental quadrupole moments for $sd$ shell nuclei have been evaluated and the recommended values were presented
in Ref. \cite{DeRydt2013} along with shell-model calculations using the USD and SDPF-U interactions.
We have used these experimental quadrupole moments in Table \ref{qad}. The values not available in this evaluation are taken
from the specified references. 

The experimental static and dynamic moments for Ne, Na, Mg and Al isotopes up to 20 neutrons, at the borders of (or inside)
the island of inversion, are reported in Refs. \cite{ne,ne1,na,mg,al,al1,30ne_be2,29na_be2,2930na_be2,30mg_be2,31mb_coulex,32mb_coulex,31al_g,32al_q,33al_q}.
 For explaining the intruder
configuration of neutron-rich nuclei ($\sim$ $N=20$), we need the $sd-pf$ model space.
 Using the SDPF-U-MIX effective interaction in the Ref. \cite{PRC90_014302}, it was shown the island of inversion
region emerges around $N=20$ and  $N=28$ for Ne to Al isotopes. The island of inversion is also known as
island of deformation, which is due to nucleon-nucleon correlations. Because of the correlation energy, we get a 
deformed ground-state band and the spherical mean field breaks. The normal-order filling of 
orbits in the case of island of inversion candidates ($^{30}$Na, $^{31}$Na, $^{31}$Mg, and
$^{33}$Al) vanishes, where $^{33}$Al is found to be at the border of the island of inversion \cite{33al_q}. The IM-SRG and CCEI $ab~initio$ effective interactions 
 contain excitations of particles within different shells ($\sim$ 13 oscillator major shells), projected to a particular  $sd$
model space.
The static and transitional quadrupole moments of nuclei lying in the island of inversion region show a drastic
enhancement of quadrupole collectivity compared to neighboring nuclei. This has been attributed to a combination of
a reduction in the $N=20$ shell gap due to the tensor part of the nucleon-nucleon monopole interaction and enhanced 
quadrupole correlations induced by neutron excitations across this reduced shell gap.  The moments of these isotopes
have been very well described using the phenomenological shell-model interactions in an enhanced $sd$-$pf$ model space where such
neutron excitations are included, as illustrated e.g. in Refs. \cite{PRC70_044307,30mg_be2,30ne_be2,33al_q}. 
The spectroscopy of other $sd$ shell isotopes, including their moments, has been very well described by shell-model calculations
in the $sd$ valence space using the phenomenological  effective interactions, such as USDB \cite{PRC78_064302}.
However, the recent $ab~initio$ calculations reproduce also very well the spectroscopy of these $sd$ shell isotopes, even
improving an accurate description of their structure. It will be interesting to see how well they reproduce
the ground-state static and dynamic moments.

\begin{table*}
  \begin{center}
   \leavevmode
    \caption{\label{mag}Comparison of the experimental magnetic dipole moments $(\mu_N)$ with the theoretical values calculated using free
    $g$ factors for $sd$ shell nuclei. The experimental data are taken from Refs. \cite{stone2005,be2}.}
 \begin{tabular}{ l l l l l l l }
  \hline\hline

Nuclei\hspace{8mm}  &  State \hspace{8mm}   & $E_x$ (keV) \hspace{8mm}     &  $\mu_{\rm expt}$ \hspace{8mm}            & $\mu_{\rm IM-SRG}$  \hspace{8mm}     & $\mu_{\rm CCEI}$ \hspace{8mm}    & $\mu_{\rm USDB}$ \hspace{8mm}       \\
\hline
$^{17}{\rm O}$   & $5/2^+$        &  0 & $-1.89379(9)$         & $-1.913$     &  $-1.913$    &     $-1.913$               \\
$^{18}{\rm O}$   & $2^+$      & 1982       &   $-0.57(3)$      &  $-1.094$     & $-1.022$      & $-0.799$                    \\
                 & $4^+$     & 3555        &   $2.5(4)$         &  $-2.455$      &  $-2.438 $    &$-2.172$   	           \\

$^{19}{\rm O}$  & $5/2^+$     &0    &   1.53195(7)            &  $-1.509$     &  $-1.518$        &  $-1.531$ 	          \\
                & $3/2^+$     & 96  &   $-0.72(9)$            &  $-0.885$      &  $-0.869$         & $-0.945$ 	          \\

$^{20}{\rm O}$ & $2^+$       &1674     &  $0.70(3)$            & $-0.921$      & $-0.926$          & $-0.716$   	           \\

$^{17}{\rm F}$ & $5/2^+$  &0  &  $+4.7213(3)$                 &  +4.793    &   +4.793          &      +4.793          \\

$^{18}{\rm F}$ & $3^+$       &937   &     $+1.77(12)$          & +1.847      &  +1.826     &$+1.872$               \\
               & $5^+$       &1121 &    +2.86(3)               &  +2.880    &  +2.880    &   $+2.880$             \\

$^{19}{\rm F}$ & $1/2^+$     &0 &  +2.628868(8)               & +2.917     &  +2.932       & +2.898             \\
               & $5/2^+$    & 197 &  $3.595(13)$             & +3.560      &  +3.611       & +3.584             \\

$^{20}{\rm F}$ & $2^+$     & 0 &  $+2.09335(9)$               & +2.171      &  +2.183      & $+2.092$                \\

$^{21}{\rm F}$ & $5/2^+$     & 0 &     3.9194(12)             & +3.393     &  +3.345           &   +3.779          \\

$^{22}{\rm F}$ & $4^+$      & 0   & (+)2.6944(4)            & +2.535    &  +2.477     &  $+2.540$            \\

$^{19}{\rm Ne}$& $1/2^+$    & 0& $-1.8846(8)$           &    $-2.060$  &    $-2.092$         &   $-2.037$            \\
               & $5/2^+$    & 238 & $-0.740(8)$             & $-0.608$    &  $-0.669$        &   $-0.673$            \\

$^{20}{\rm Ne}$& $2^+$  &1634  & $+1.08(8)$               &  +1.036     & +1.037      &  $+1.020$                 \\
               & $4^+$ &  4247& $+1.5(3)$                &  +2.086     & +2.095      &  $+2.052$          \\

$^{21}{\rm Ne}$& $3/2^+$ &  0     &  $-0.661797(5)$          & $-0.665$      &  $-0.586$    & $-0.750$              \\
               & $5/2^+$ &  351   &  0.49(4)                & $-0.350$      &  $-0.365$    &  $-0.574$            \\

$^{22}{\rm Ne}$& $2^+$ &1275    & $+0.65(2)$              &  +0.616     &   +0.550   & $+0.748$            \\
               & $4^+$ & 3357   &$+2.2(6)$              &  +1.623       &  +1.332   & $+2.044$          \\

$^{23}{\rm Ne}$& $5/2^+$ &  0 &$-1.077(4)$            & $-0.854$        &   $-0.786$    & $-1.050$            \\

$^{25}{\rm Ne}$& $1/2^+$   &  0   &  $-1.0062(5)$     &  $-0.657$      &$-0.924$    &$-0.928$       \\

$^{20}{\rm Na}$& $2^+$ & 0 &$+0.3694(2)$              & +0.390    &  +0.330    &  $+0.446$                \\

$^{21}{\rm Na}$& $3/2^+$ & 0& +2.83630(10)          &  +2.445     &  +2.388          &  $+2.489$ \\
              &  $5/2^+$ & 332 & $3.7(3)$               & +3.194       & +3.196           & $+3.355$ \\

$^{22}{\rm Na}$& $3^+$ & 0  & $+1.746(3)$              & +1.798     & $+1.806$   &  $+1.791$              \\
               & $1^+$ & 583  &$+0.523(11)$             & +0.506         &$+0.529$    & $+0.518$               \\
$^{23}{\rm Na}$& $3/2^+$ & 0&$+2.2176556(6)$         & +1.972       & +1.887           & $+2.098$     \\
$^{24}{\rm Na}$& $4^+$ &  0&  $+1.6903(8)$           & +1.377   & $+1.285$    & $+1.631$                \\
               & $1^+$ & 472 &  $-1.931(3)$            & +0.908  & $-0.881$    & $-1.865$                \\
$^{25}{\rm Na}$& $5/2^+$ & 0 & $+3.683(4)$           & +2.934       &  +3.361            &$+3.367$         \\

$^{26}{\rm Na}$& $3^+$ & 0& $+2.851(2)$             & +2.296         & $+2.360$      & $+2.632$              \\

$^{27}{\rm Na}$& $5/2^+ $& 0 &   $+3.895(5)$        & +3.230        &  +3.623           & $+3.647$          \\

$^{28}{\rm Na}$& $1^+$ & 0 & $+2.426(5)$             & +2.146         & $+1.760$   & $+2.081$        \\

$^{29}{\rm Na}$& $3/2^+$ & 0& $+2.449(8)$           &  +2.181       &  +2.198         &  $+2.438$       \\

$^{30}{\rm Na}$& $2^+$ & 0& $+2.083(10)$             & +2.245          & $+2.883$    &  $+2.418$     \\

$^{31}{\rm Na}$& $ 3/2^+ $ & 0    & $+2.305(8)$       &  +2.535       &  +2.551         & $+2.614$       \\
$^{21}{\rm Mg}$& $5/2^+$ & 0 & $-0.983(7)$        &  $-0.342$       &  $-0.351$             & $-0.848$    \\

$^{23}{\rm Mg}$& $3/2^+$ & 0 & $-0.5364(3)$        &  $-0.305$       &  $-0.218$             & $-0.410$ \\

$^{24}{\rm Mg}$& $2^+$   & 1369 & $+1.076(26)$       &  +1.050       &  $+1.094$     &  $+1.026$      \\
               & $4^+$  & 4123 & $+1.6(12)$       &  +2.103        &  $+2.169$      & $+2.070$        \\
               & $2^+$   & 4238 & $+1.2(4)$        &  +1.072        & $+1.062$        & $+1.037$       \\
               & $4^+$  & 6010 & $+2.0(16)$       &  +2.095        &  $+2.115$       &  $+2.048$       \\

$^{25}{\rm Mg}$& $5/2^+$ & 0& $-0.85545(8)$      &   $-0.617$        & $-0.197$         &$-0.849$           \\

$^{26}{\rm Mg}$& $2^+$ & 1809& $+1.0(3)$            &  +1.024         & $+1.281$        &  $+1.739$               \\

$^{27}{\rm Mg}$& $1/2^+$ & 0 & $-0.411(2)$      &  +0.197         & $-0.256$               & $-0.412$       \\

\hline
\hline
\end{tabular}
\end{center}
\end{table*}
\addtocounter{table}{-1}
\begin{table*}
  \begin{center}
    \leavevmode
    \caption{{\label{mag} Continuation.\/}}
\begin{tabular}{l l l l l l l l}
\hline\hline

Nuclei\hspace{8mm}  &  State \hspace{8mm}   & $E_x$ (keV) \hspace{8mm}     &  $\mu_{expt}$ \hspace{8mm}            & $\mu_{\rm IM-SRG}$  \hspace{8mm}     & $\mu_{\rm CCEI}$ \hspace{8mm}    & $\mu_{\rm USDB}$ \hspace{8mm}       \\
\hline

$^{29}{\rm Mg}$& $3/2^+$  &  0  & $+0.9780(6)$   &   +1.114       &  +1.470              & $+1.071$       \\

$^{31}{\rm Mg}$& $1/2^+$ & 0  & $-0.88355(15)$   &  $-0.563$         &   +1.406            &  $-0.923$    \\

$^{23}{\rm Al}$& $5/2^+$ & 0 & $+3.889(5)$    &   +3.716         &     +3.681        &  +3.866    \\

$^{24}{\rm Al}$& $1^+$ & 426 & $2.99(9)$    &   +2.660\footnote{Here we have reported the shell model result of second $1^+$ state, because calculated order of levels are $1^+-4^+-1^+$, while the experimental g.s. is $4^+$.}        &  +2.071           & +2.985     \\

$^{25}{\rm Al}$& $5/2^+$ & 0 & $3.6455(12)$    &  +3.462         &  +3.142            & +3.655     \\

$^{26}{\rm Al}$& $5^+$ & 0 & $+2.804(4)$    &   +2.850        &    +2.907          & +2.839    \\

$^{27}{\rm Al}$& $5/2^+$ & 0  & $+3.6415069(7)$   &   +2.525        &  +2.461            & +3.455     \\

$^{28}{\rm Al}$& $3^+$ & 0  & $3.242(5)$   &   +2.718        &   +2.378           & +3.098    \\
               & $2^+$  & 31 & $+4.3(4)$   &   +1.044        &   +0.675           & +3.215     \\
               
$^{30}{\rm Al}$& $3^+$ & 0 & $3.010(7)$    &  +2.442          &   +3.455          &   +3.039   \\               

$^{31}{\rm Al}$& $(5/2^+)$ & 0  & $+3.830(5)$   &  +3.571         & +3.863             & +3.761     \\

$^{32}{\rm Al}$& $1^+$ & 0 & $ 1.952(2)$    &   +1.485        &   +1.811          &   +1.612   \\

$^{33}{\rm Al}$& $(5/2^+)$ & 0 & $ +4.088(5)$    &    +4.012       &    +4.268         &  +4.224    \\

$^{27}{\rm Si}$& $5/2^+$ & 0  & $(-)0.8554(4)$   &   +0.117        &  +0.337            & $-0.678$    \\

$^{28}{\rm Si}$& $2^+$ & 1779 & $+1.1(2)$     &   +1.040        &  +1.093            & +1.031     \\

$^{29}{\rm Si}$& $1/2^+$ & 0 & $-0.55529(3)$    &    $-0.010$       &  $-0.575$            & $-0.503$     \\

$^{30}{\rm Si}$& $2^+$ &  2235 & $+0.8(2)$    &   +0.839        &  +1.939            & +0.732     \\

$^{33}{\rm Si}$& $(3/2^+)$ & 0 & $1.21(3)$    & +1.212          &  +1.803            & +1.206      \\

$^{28}{\rm P}$& $3^+$ & 0 & $0.312(3)$    &   +1.648      &    $+1.076$      &   $+0.302$  \\

$^{29}{\rm P}$& $1/2^+$ & 0 & $1.2346(3)$    &  +0.558        &  $+1.348$            & $+1.133$     \\

$^{31}{\rm P}$& $1/2^+$ & 0 & $+1.13160(3)$    &  $+0.081$        &  $+1.694$            & $+1.087$     \\
             & $3/2^+$ & 1270 & $+0.30(8)$    & $+0.318$          &  $-0.063$            & $+0.167$     \\
              & $5/2^+$ & 2230 & $+2.8(5)$    &  $+1.260$         &  $+3.097$            &  $+2.218$    \\

$^{32}{\rm P}$& $1^+$ & 0& $-0.2524(3)$    &  $-0.764$         &  $+0.177$            & $-0.021$     \\

$^{31}{\rm S}$& $1/2^+$ & 0& $0.48793(8)$    &  $+0.472$      &  $-1.003$            & $-0.441$    \\

$^{32}{\rm S}$& $2^+$ & 2231 & $+0.9(2)$    &   $+1.022$        &   $+0.980$           &  $+1.010$   \\
              & $4^+$ & 4459 & $+1.6(6)$    &   $+2.046$        &    $+1.840$          & $+2.028$    \\
  \hline\hline
\end{tabular}
\end{center}
\end{table*}

\begin{table*}
  \begin{center}
    \leavevmode
    \caption{\label{qad}Comparison of the experimental quadrupole moments ($eb$) with the theoretical values calculated by using
    $e_p$=1.5e and $e_n$=0.5e.}
 \begin{tabular}{ l l l l l l l l}
  \hline\hline
Nuclei \hspace{8mm}       &  State \hspace{8mm}   & $E_x$ (keV)  & $Q_{\rm expt}$ \hspace{8mm}    & $Q_{\rm IM-SRG}$ \hspace{8mm}   & $Q_{\rm CCEI}$\hspace{8mm}  &$Q_{\rm USDB}$ \hspace{8mm}     & Ref.      \\
\hline

$^{17}{\rm O}$ & $5/2^+$  &  0   & $-0.02558(22)$       &  $-0.0302$     & $-0.0302$       &    $-0.0302$    &  \cite{Ne21}     \\

$^{18}{\rm O}$ & $2^+$  &1982 & $-0.036(9)$             &  $-0.0153$       &   $-0.0172$ &  $-0.0294$  &  \cite{stone1} \\

$^{19}{\rm O}$ & $5/2^+$ & 0 & $0.00362(13)$            &  $+0.0003$      & $+0.0005$         & $-0.0026$   & \cite{DeRydt2013}              \\

$^{17}{\rm F}$ & $5/2^+$ & 0&  $0.0799(34)$              & $-0.0907$      &   $-0.0907$            &   $-0.0907$      &  \cite{DeRydt2013}         \\

$^{18}{\rm F}$ & $5^+$ & 1121  & $0.077(5)$              & $-0.1226$     &  $-0.1226$    	&$-0.1224$	&\cite{stone1}            \\
$^{19}{\rm F}$ & $5/2^+$ & 197 &  $0.0942(9)$           & $-0.1048$	 & $-0.01056$	&$-0.1045$  & \cite{F19} \\

$^{20}{\rm F}$ & $2^+$ &0  & $0.0547(18)$              &  $+0.0677$     &$+0.0729$ 	&$+0.0679$ 	 & \cite{DeRydt2013}           \\

$^{21}{\rm F}$ & $5/2^+$ &0 &   $0.0943(33)$           & $-0.1180$	& $-0.1175$	&$-0.1199$  & \cite{DeRydt2013}  \\

$^{22}{\rm F}$ & $4^+$  &0  & $0.003(2)$               &  $-0.0167$       &$-0.0249$ 	&$-0.0078$  & \cite{stone1}	            \\

$^{20}{\rm Ne}$& $2^+$ &1634 & $-0.23(3)$             &  $-0.1573$     &$-0.1578$ 	&$-0.1576$  & \cite{stone1}	          \\

$^{21}{\rm Ne}$& $3/2^+$& 0&  $+0.10155(75)$          &  $+0.1127$      & $+0.1109$	& $+0.1119$    & \cite{DeRydt2013,Ne21}      \\

$^{22}{\rm Ne}$& $2^+$ & 1257 & $-0.19(4)$             &  $-0.1561$	&$-0.1536$ 	&$-0.1532$ & \cite{stone1}	         \\

$^{23}{\rm Ne}$& $5/2^+$ & 0& $+0.1429(43)$           & $+0.1728$         & $+0.1699$	&  $+0.1629$    & \cite{DeRydt2013}      \\

$^{20}{\rm Na}$& $2^+$  & 0& $0.1009(88)$   	      &  $+0.0961$ & $+0.100$	&     $+0.0946$	     & \cite{DeRydt2013}      \\
$^{21}{\rm Na}$& $3/2^+$ &0 & $0.137(12)$       	 & $+0.1224$	& $+0.1216$		& +0.1218     & \cite{DeRydt2013}      \\

$^{22}{\rm Na}$& $3^+$ &0 & $+0.167(17)$              &  $+0.2496$ &$+0.2405$       &$+0.2506$	 & \cite{DeRydt2013}	          \\

$^{23}{\rm Na}$& $3/2^+$ & 0 & $+0.104(1)$            & $+0.1217$        &  $+0.1246$              & $+0.1180$    & \cite{DeRydt2013,Na23}      \\

$^{25}{\rm Na}$& $5/2^+$ &0 &$0.00146(22)$            & $+0.0214$        &  $+0.0674$               & $+0.0025$   & \cite{DeRydt2013}      \\
$^{26}{\rm Na}$& $3^+$ &0 & $0.00521(20)$           & $-0.0056$        &   $+0.0239$     &$-0.0051$      & \cite{DeRydt2013}     \\

$^{27}{\rm Na}$& $5/2^+$ &0 & $0.00708(24)$           & $-0.0120$        & $-0.0035$         &$-0.0127$     & \cite{DeRydt2013}      \\

$^{28}{\rm Na}$& $1^+$ & 0& $0.0389(11)$         	 & $+0.0539$          &$+0.0368$      & $+0.0495$    & \cite{DeRydt2013}           \\

$^{29}{\rm Na}$& $3/2^+$ &0 & $+0.0842(25)$      	   & $+0.0737$        & $+0.1046$         &$+0.0791$     & \cite{DeRydt2013}      \\

$^{30}{\rm Na}$& $2^+$ &0 & $+0.146(1.6)$      	   & $-0.1122$        & $-0.1048$         &     $-0.1149$   & \cite{epja}   \\

$^{31}{\rm Na}$& $3/2^+$ &0 & $+0.105(2.5)$      &  $+0.0465$       & $+0.0920$         & $+0.0583$      & \cite{epja}      \\

$^{23}{\rm Mg}$& $3/2^+$ &0  & $0.1133(37)$       	  & $+0.1285$        &  $+0.1322$               & $+0.1229$    & \cite{DeRydt2013}      \\

$^{24}{\rm Mg}$& $2^+$ &1369  & $-0.29(3)$          & $-0.1914$        &  $-0.1857$       & $-0.1931$         & \cite{stone1}      \\
$^{25}{\rm Mg}$& $5/2^+$ & 0 & $+0.1994(20)$        & $+0.2235$        &  $+0.1809$                &  $+0.2243$     & \cite{Mg25}   \\

$^{26}{\rm Mg}$& $2^+$ & 1809& $-0.21(2)$            & $-0.1747$        & $+0.1155$       &  $-0.1439$   & \cite{stone1}     \\

$^{25}{\rm Al}$& $5/2^+$& 0 & $ 0.249(18)$         &   $+0.1949$                 &$+0.1813$    &   $+0.2018$    & \cite{DeRydt2013}    \\

$^{26}{\rm Al}$& $5^+$ & 0& $+0.259(29)$    		&   $+0.3260$        &   $+0.294$           & $+0.3028$    & \cite{DeRydt2013}  \\

$^{27}{\rm Al}$& $5/2^+$ & 0 & $+0.1466(10)$  	     & $+0.1563$          &  $+0.091$            & $+0.1803$    & \cite{DeRydt2013,Al27} \\

$^{28}{\rm Al}$& $3^+$ & 0 & $0.172(12)$   		 &   $+0.2289$        &  $+0.1388$            & $+0.1877$   & \cite{DeRydt2013}   \\

$^{31}{\rm Al}$& $5/2^+$ & 0 & $0.1365(23)$   		 &  $+0.1836$         &  $+0.1320$          &  $+0.1706$   & \cite{33al_q}\\

$^{32}{\rm Al}$& $1^+$ & 0 & $0.0250(21)$   		 &  $+0.0370$         &  $+0.006$           &  $+0.0310$   & \cite{DeRydt2013}  \\

$^{33}{\rm Al}$& $5/2^+$ & 0 & $0.141(3)$ 		 &   $+0.1375$        &  $+0.1368$            &  $+0.1390$   & \cite{33al_q}   \\

$^{27}{\rm Si}$& $5/2^+$ & 0 & $0.063(14)$  	        & $+0.1291$          & $+0.072$             & $+0.1409$    & \cite{Si27} \\

$^{28}{\rm Si}$& $2^+$ & 1779 & $+0.16(3)$    		&  $+0.2332$         &  $+0.196$            & $+0.2087$    & \cite{stone1}  \\

$^{30}{\rm Si}$& $2^+$ & 2235 & $-0.05(6)$   		 &  $+0.0465$         &  $+0.1470$            & $+0.0239$    & \cite{stone1}  \\

$^{32}{\rm S}$& $2^+$ & 1941 & $-0.15(2)$   		 &     $-0.0140$      &   $-0.0801$           & $-0.1283$    & \cite{stone1}  \\

$^{33}{\rm S}$& $3/2^+$ & 0 & $-0.0678(13)$   		 &   $-0.1431$          &   $-0.0565$           &   $-0.0736$  & \cite{S33} \\

\hline\hline
\end{tabular}
\end{center}
\end{table*}

\begin{figure*}
\centering
\includegraphics[width=6.2cm,height=5.5cm,clip]{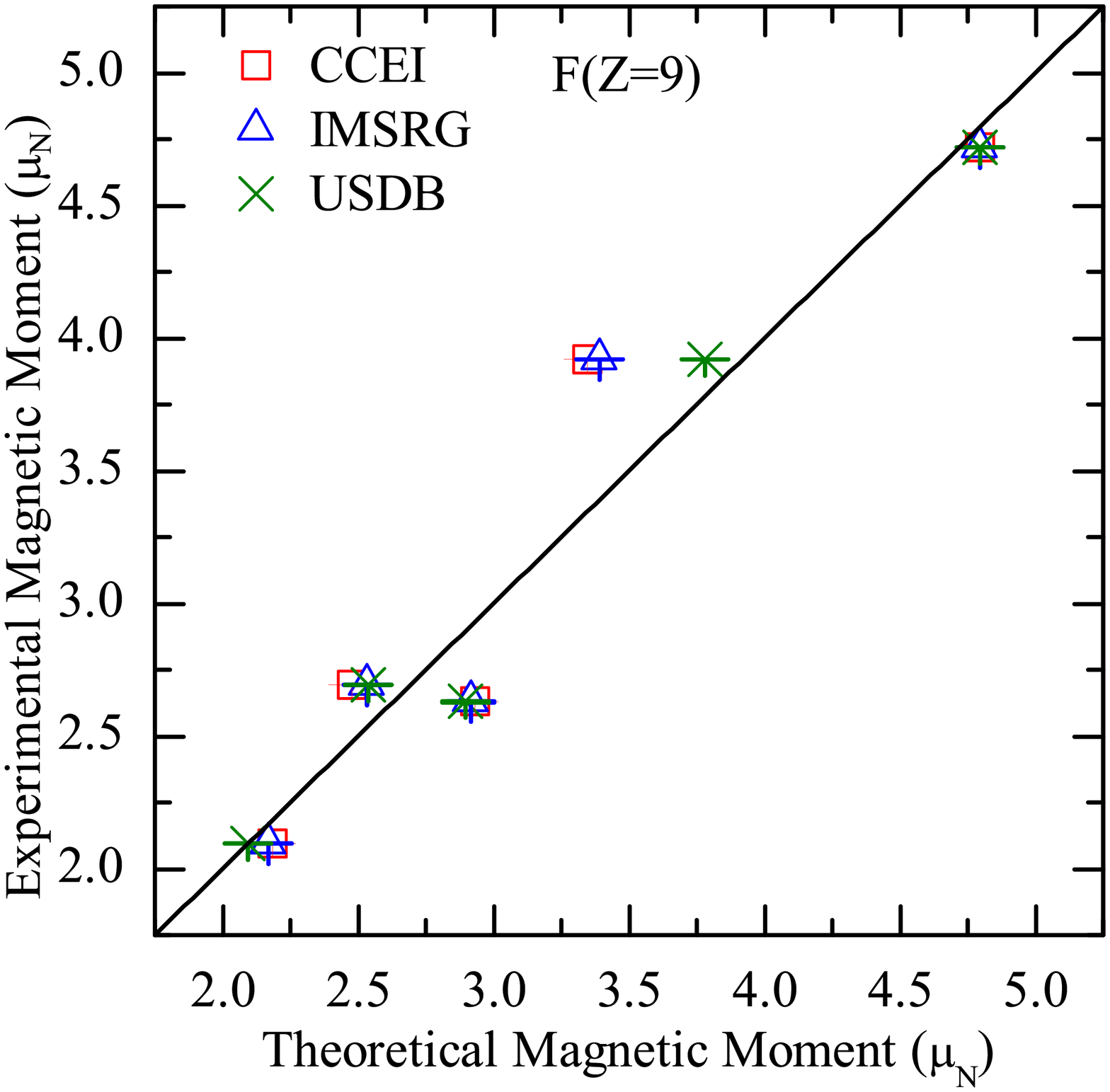}
\includegraphics[width=6.2cm,height=5.5cm,clip]{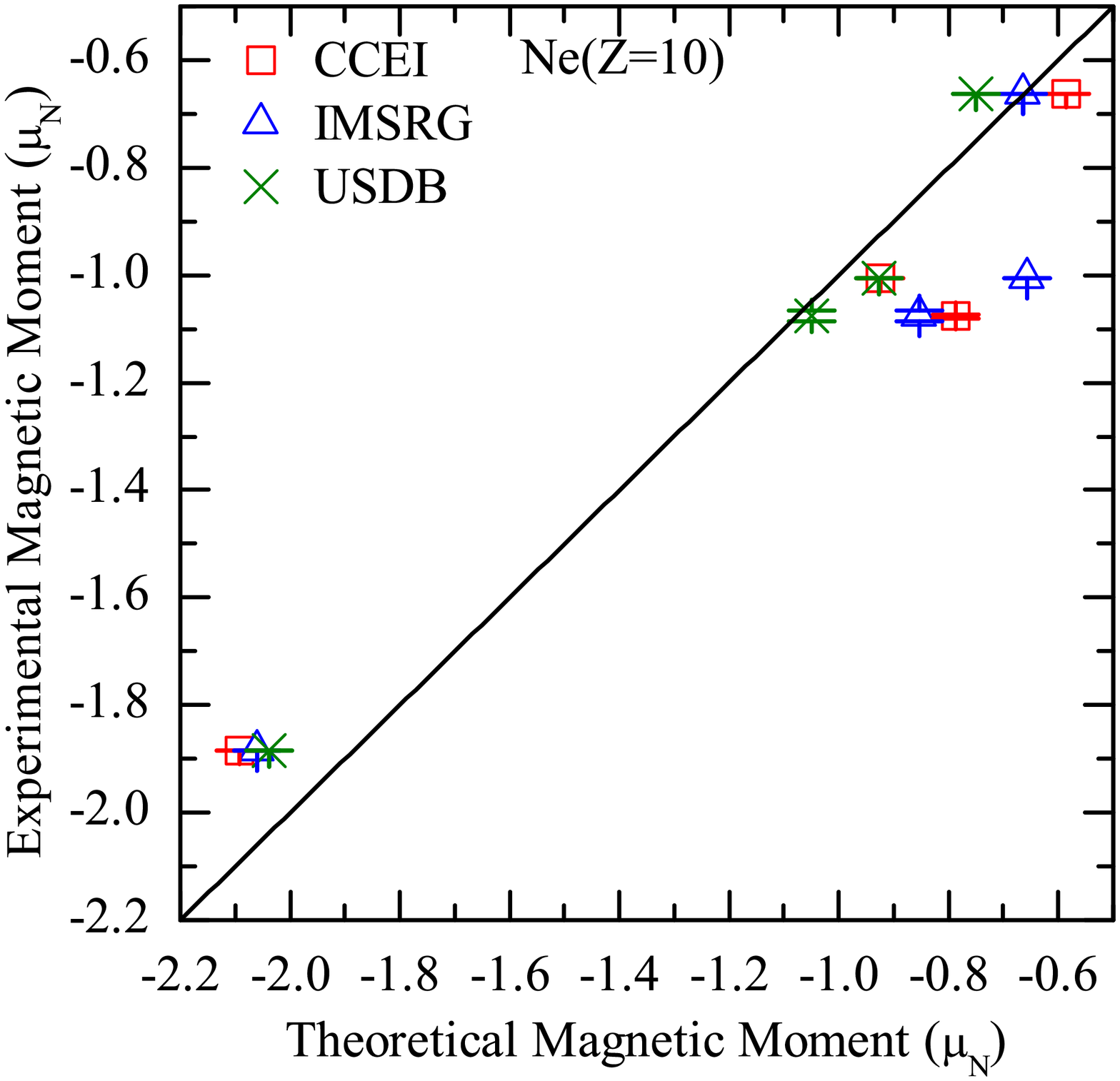}
\includegraphics[width=6.2cm,height=5.5cm,clip]{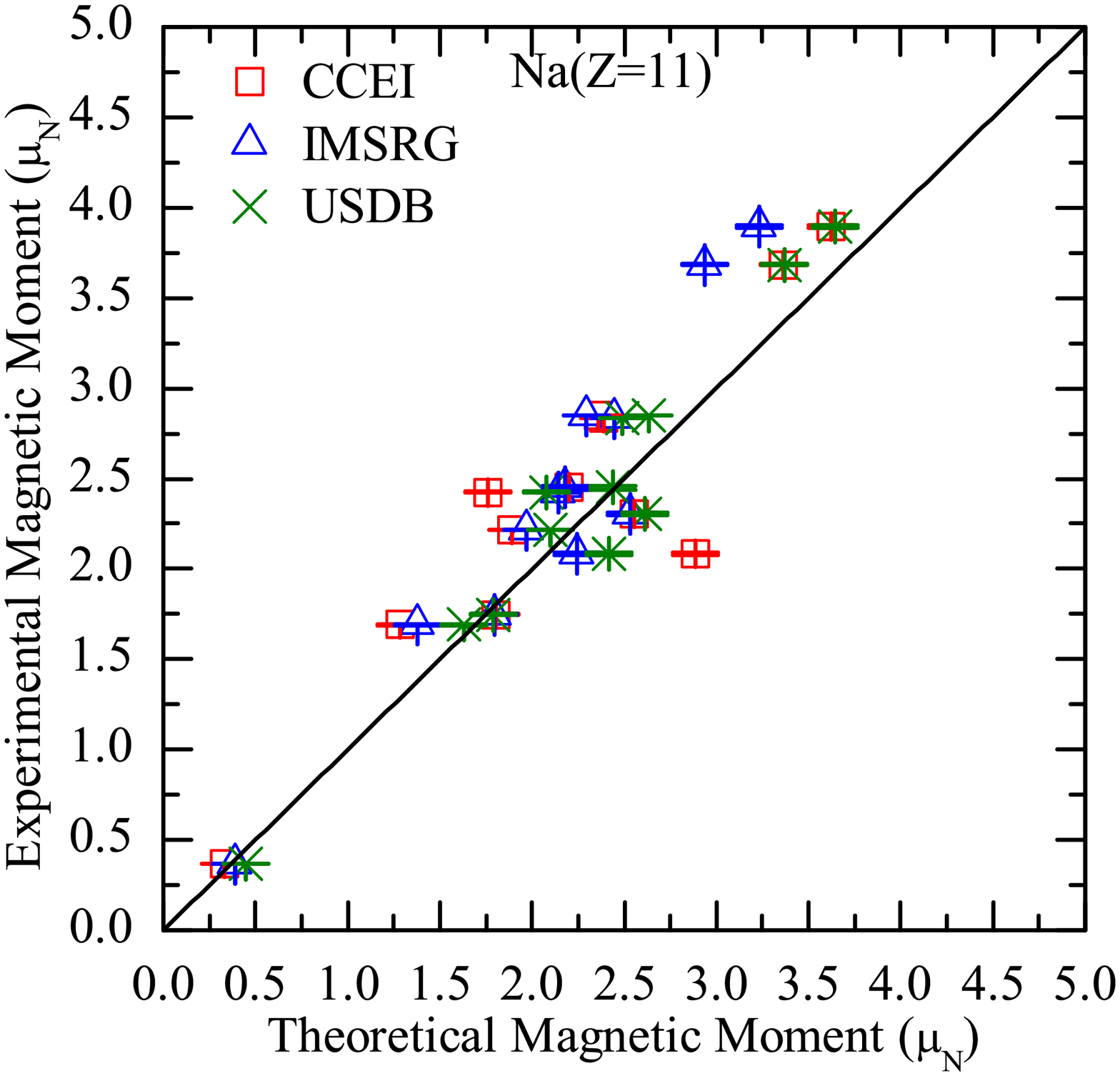}
\includegraphics[width=6.2cm,height=5.5cm,clip]{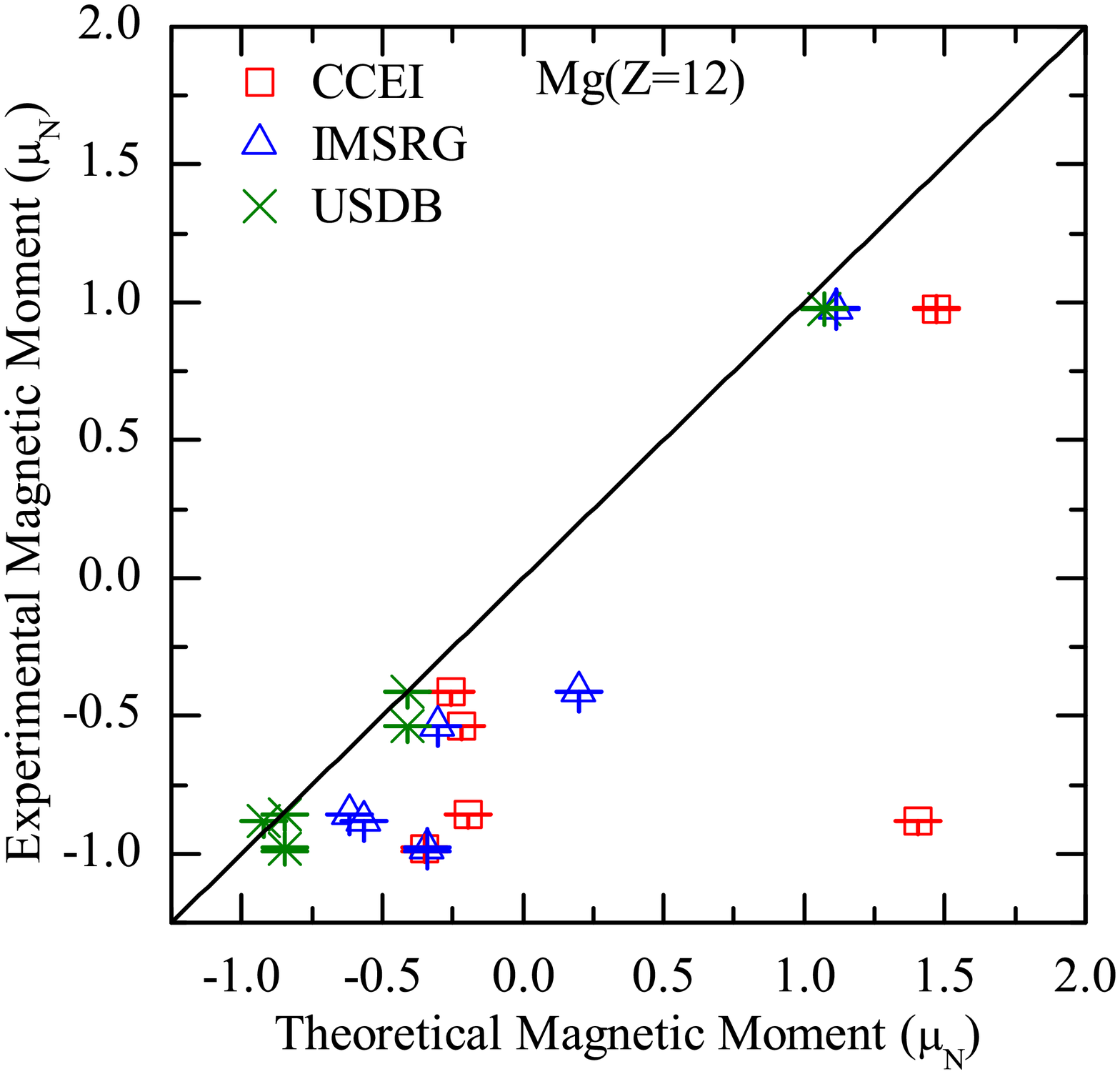}
\includegraphics[width=6.2cm,height=5.5cm,clip]{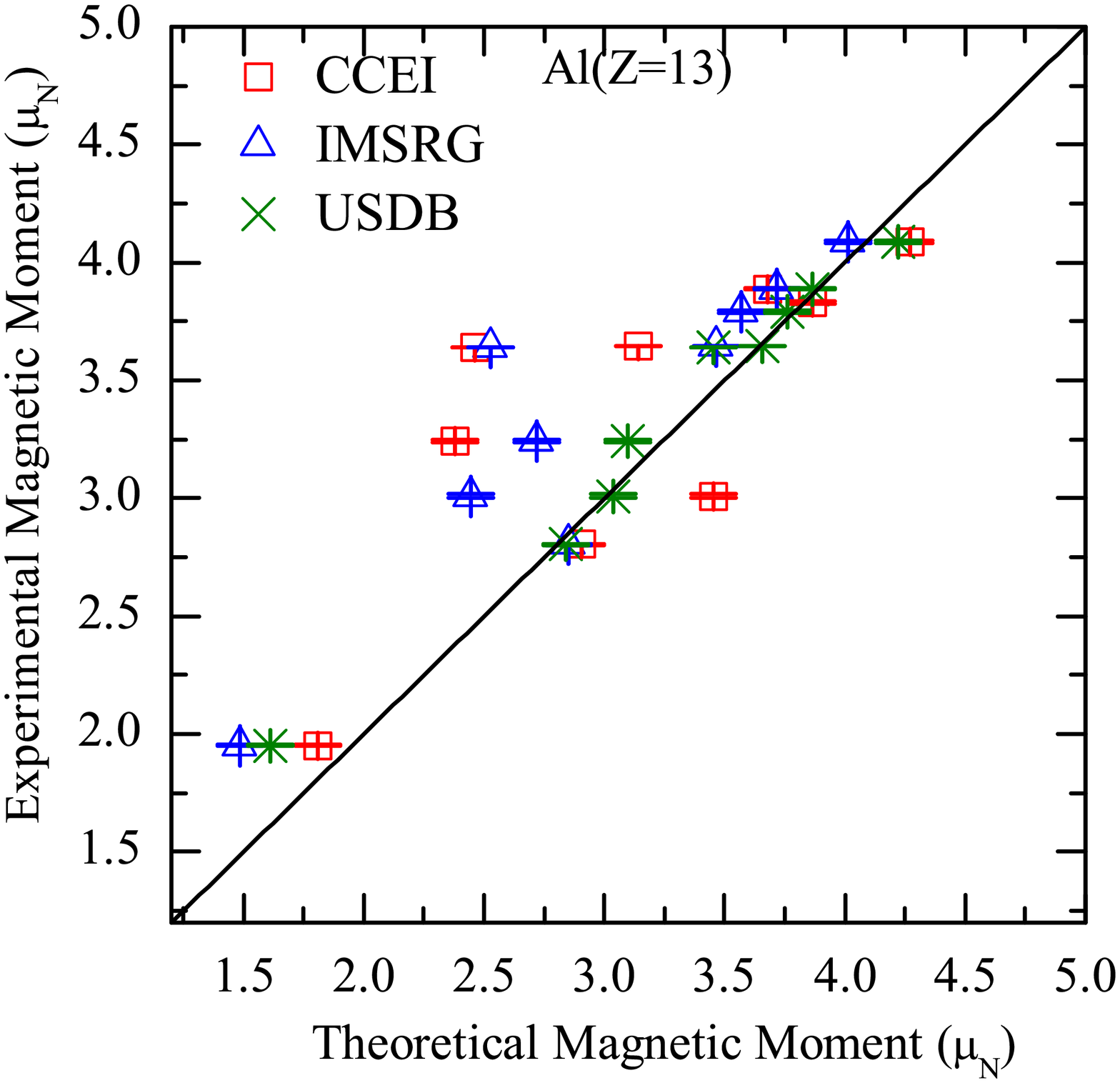}
\includegraphics[width=6.2cm,height=5.5cm,clip]{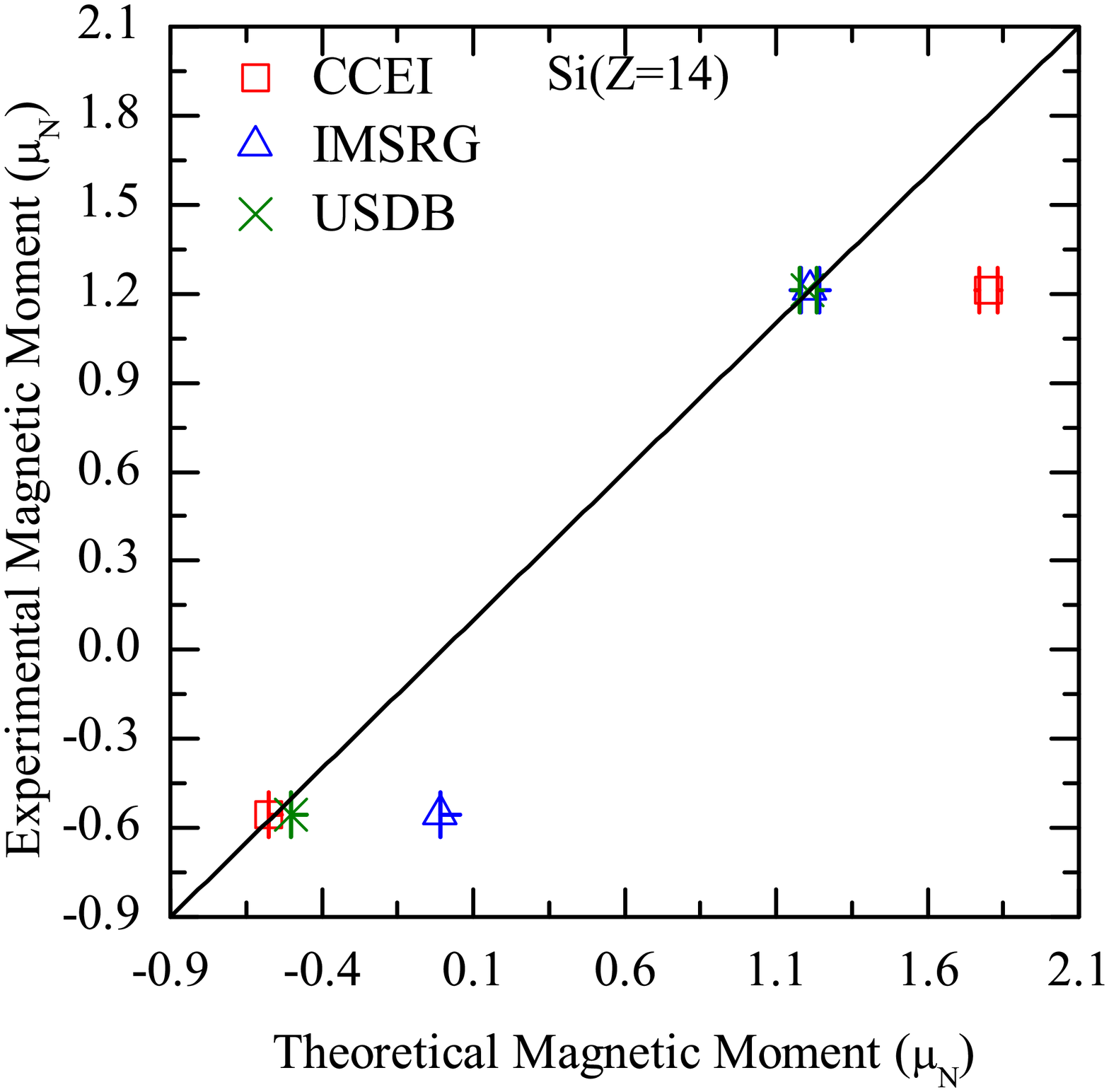}
\includegraphics[width=6.2cm,height=5.5cm,clip]{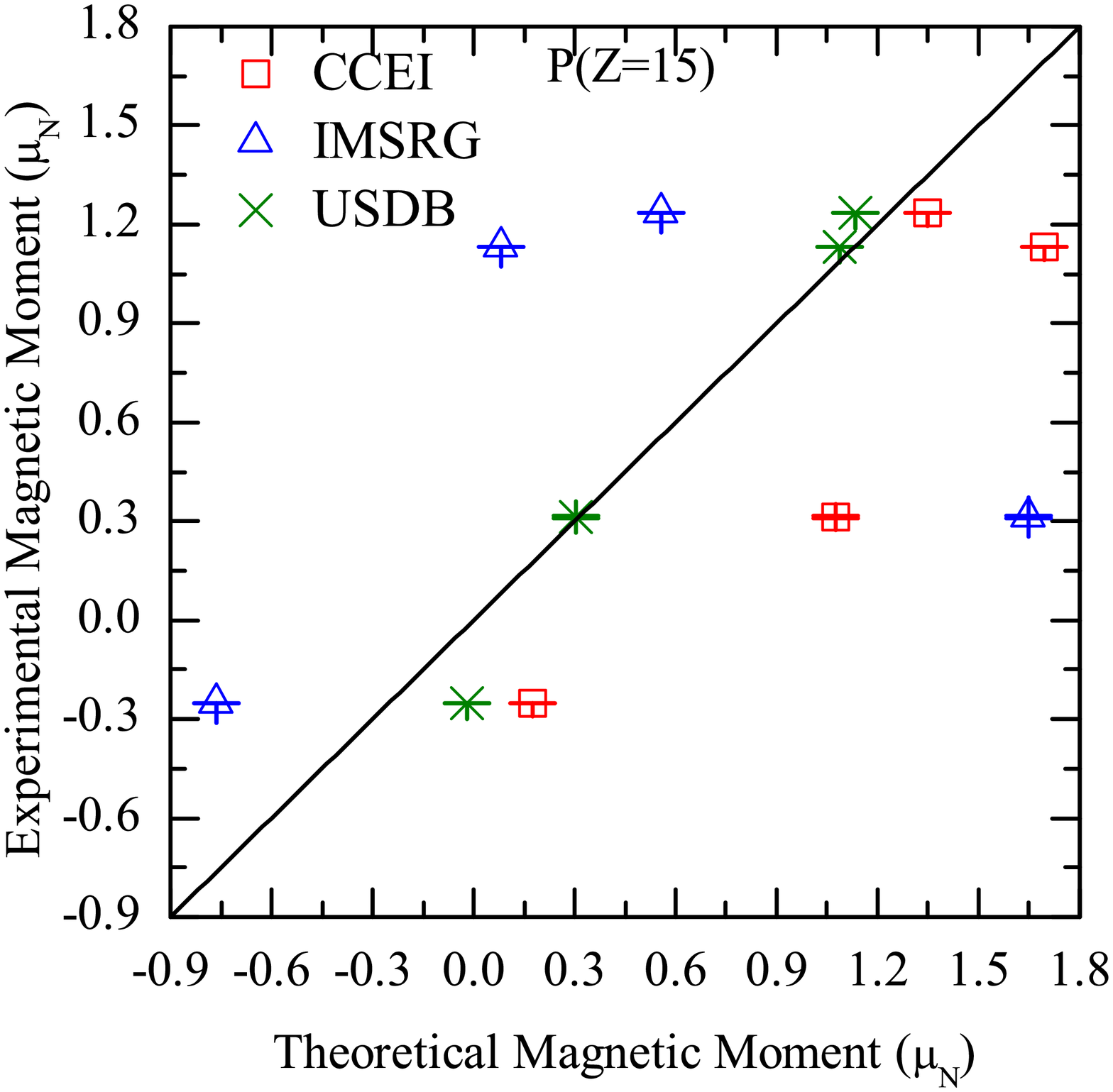}
\caption { {\label{Mag}Comparison between the experimental and theoretical magnetic dipole moments for F, Ne, Na, Mg, Al, Si, and P isotopes.
The calculated shell model signs are used in the cases when it was not measured.}}
\end{figure*}
\begin{figure*}
\centering
\includegraphics[width=7cm,height=6.5cm,clip]{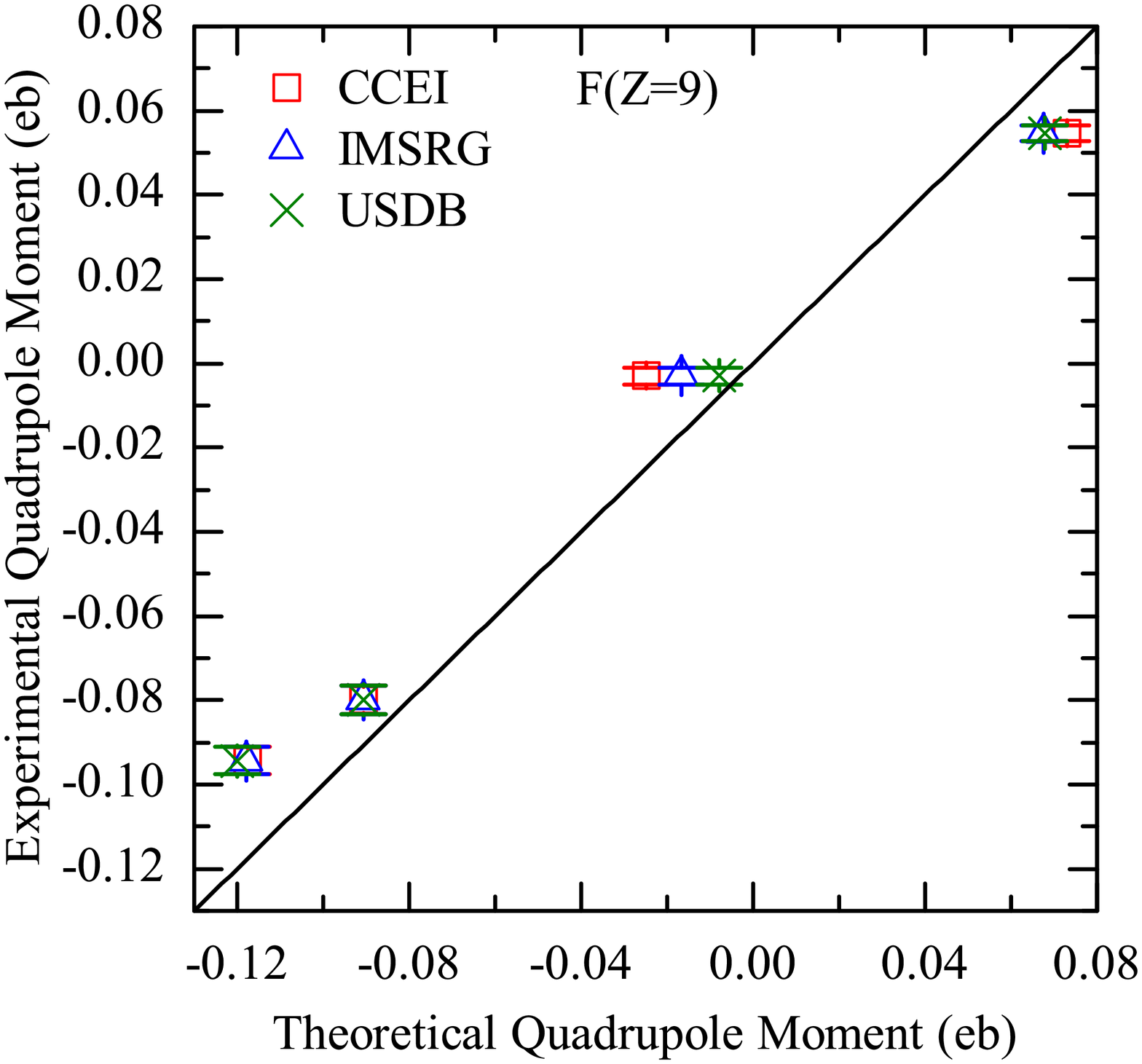}
\includegraphics[width=7cm,height=6.5cm,clip]{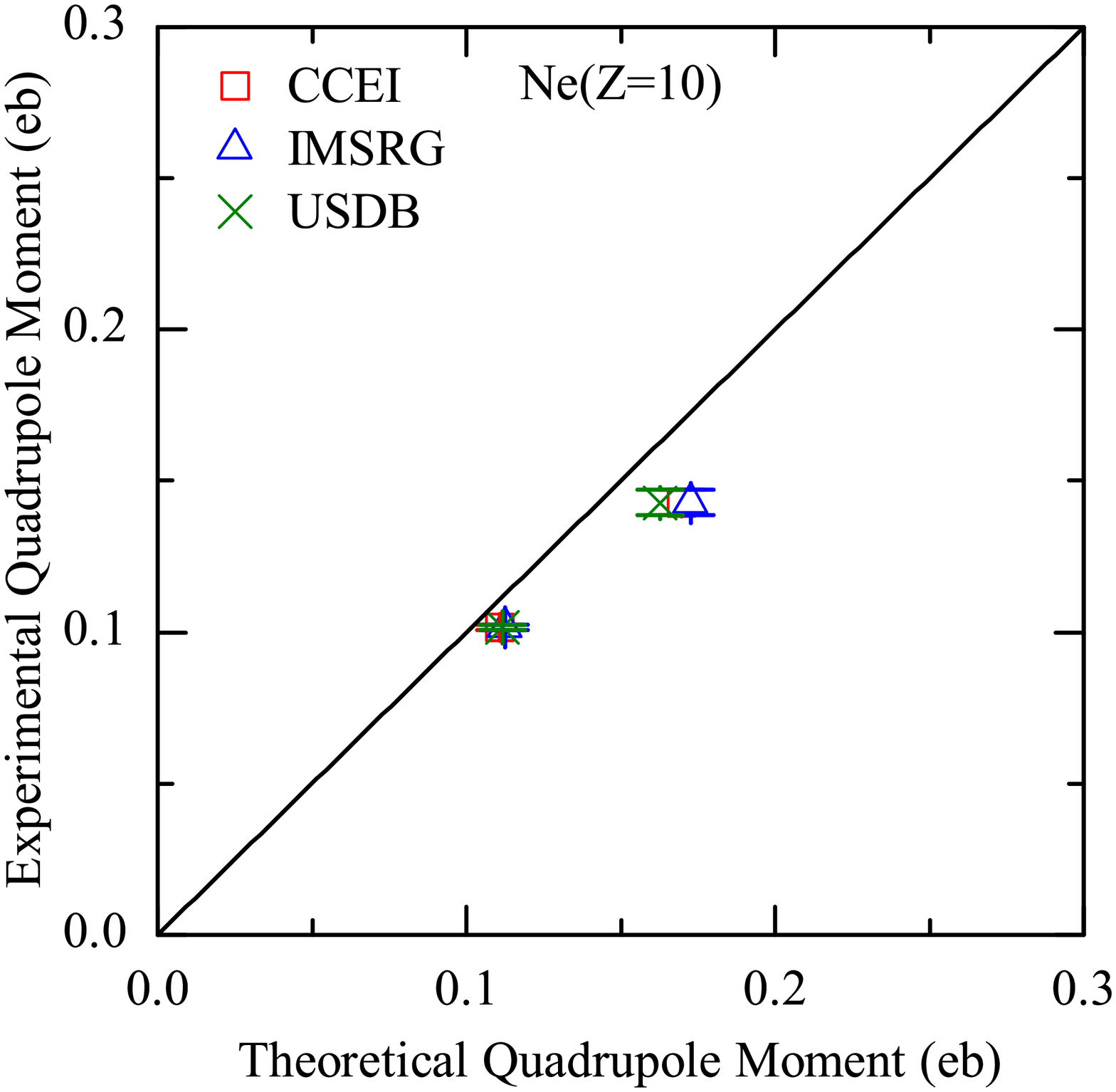}
\includegraphics[width=7cm,height=6.5cm,clip]{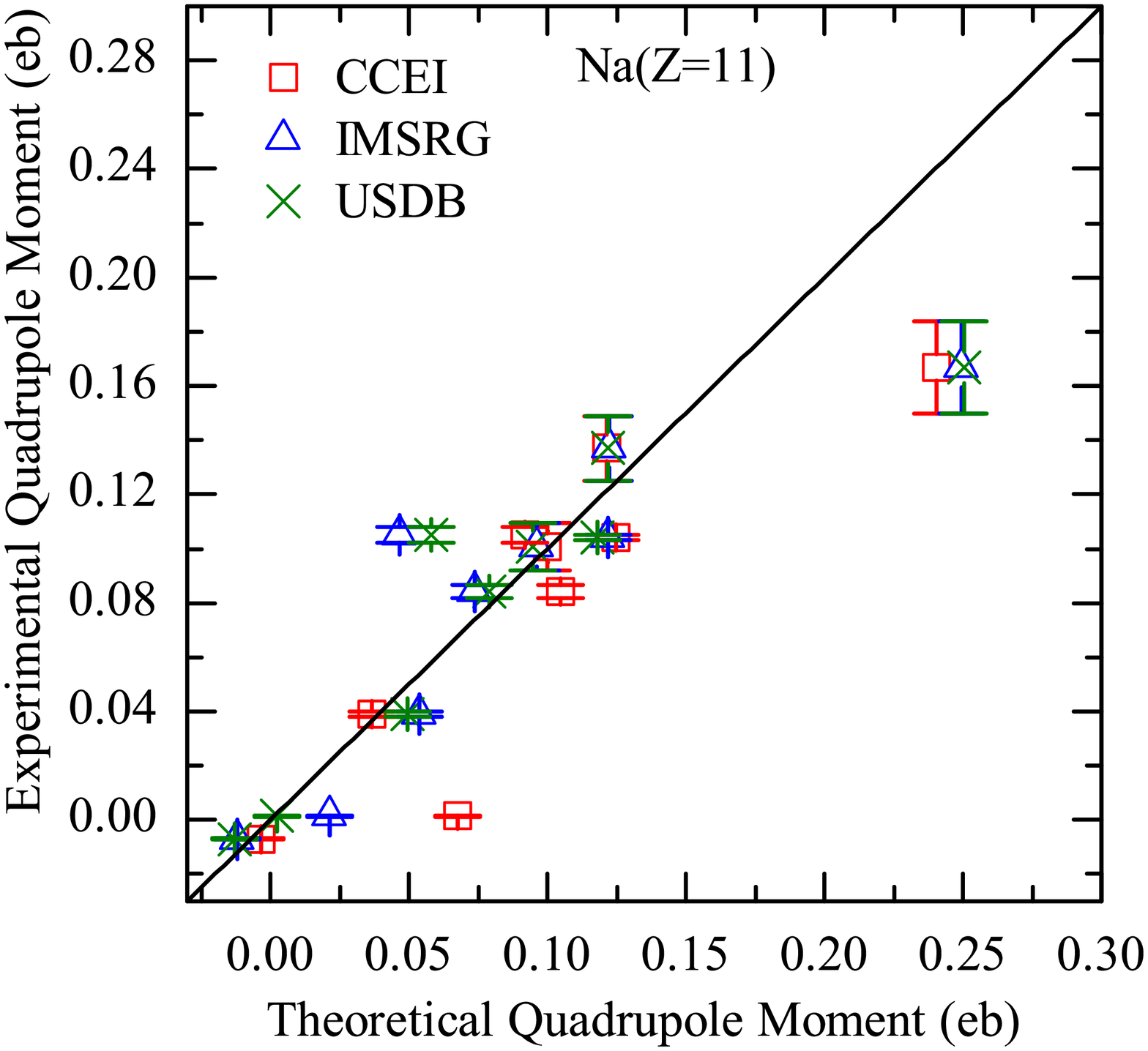}
\includegraphics[width=7cm,height=6.5cm,clip]{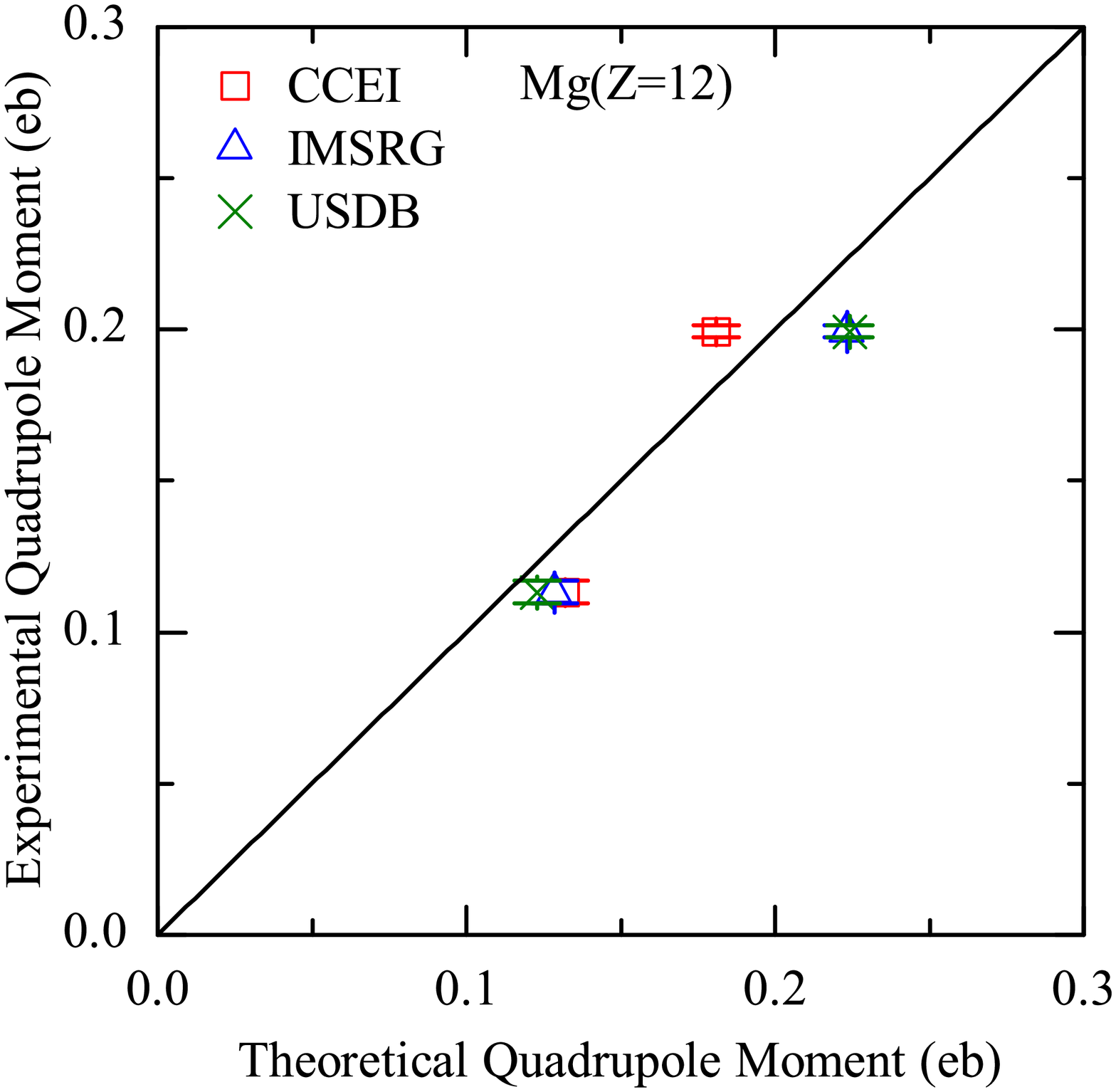}
\includegraphics[width=7cm,height=6.5cm,clip]{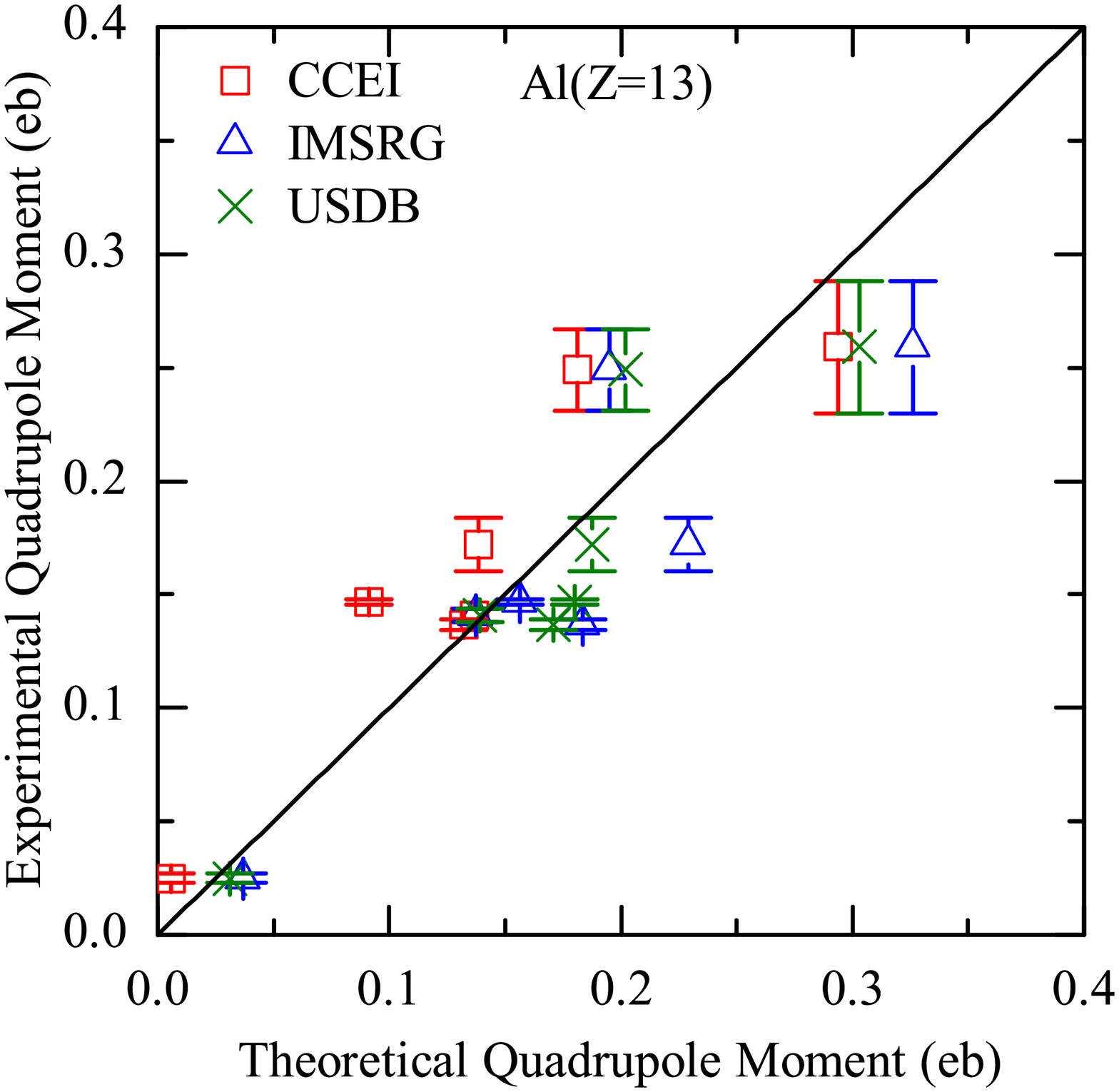}
\caption {{\label{Quad} Comparison between experimental and theoretical quadrupole moments for F, Ne, Na, Mg and Al isotopes.
The calculated shell model signs are used in case when it was not measured.
}}

\end{figure*}

\begin{figure*}
\centering
\includegraphics[width=8cm,height=6cm,clip]{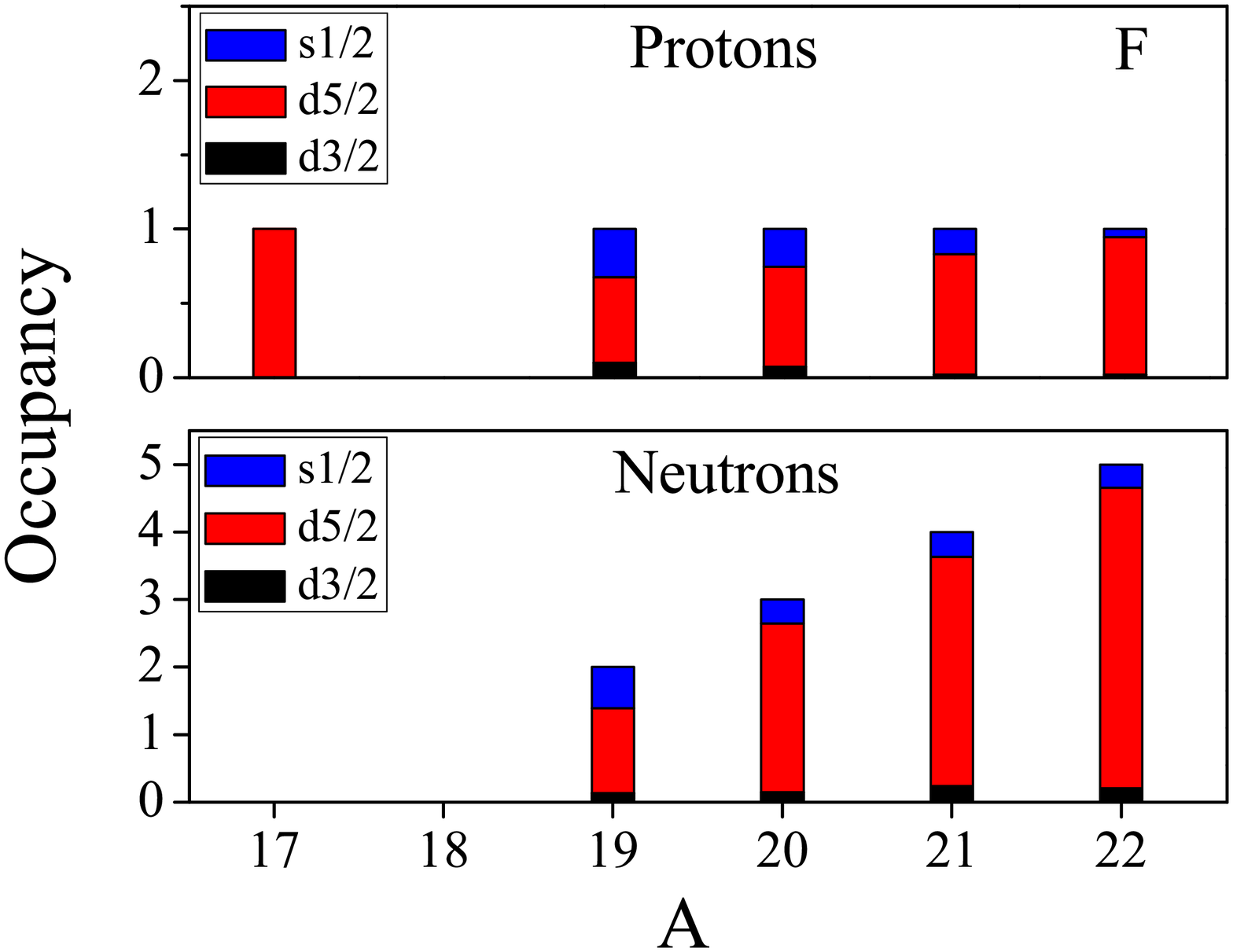}
\includegraphics[width=8cm,height=6cm,clip]{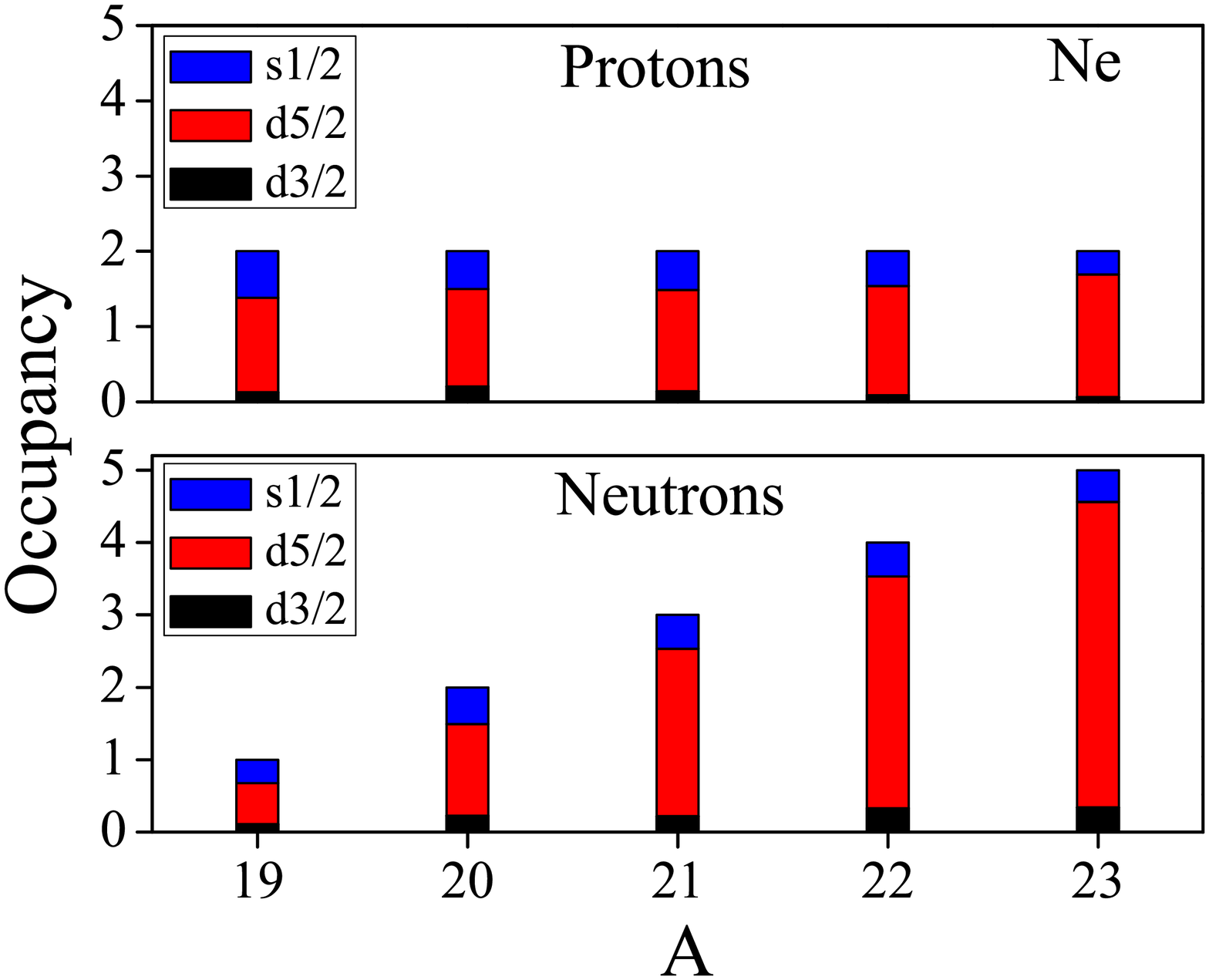}
\includegraphics[width=8cm,height=6cm,clip]{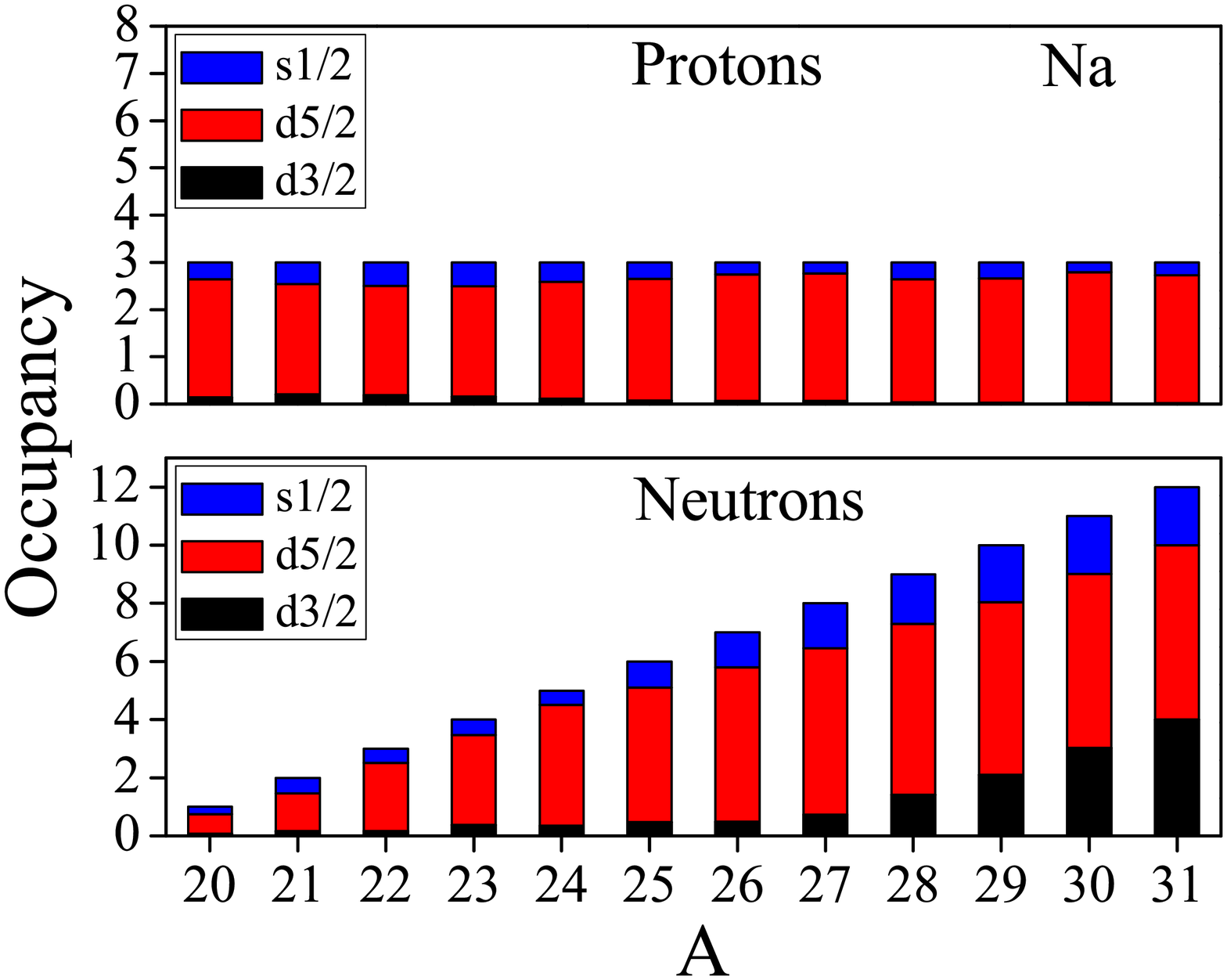}
\includegraphics[width=8cm,height=6cm,clip]{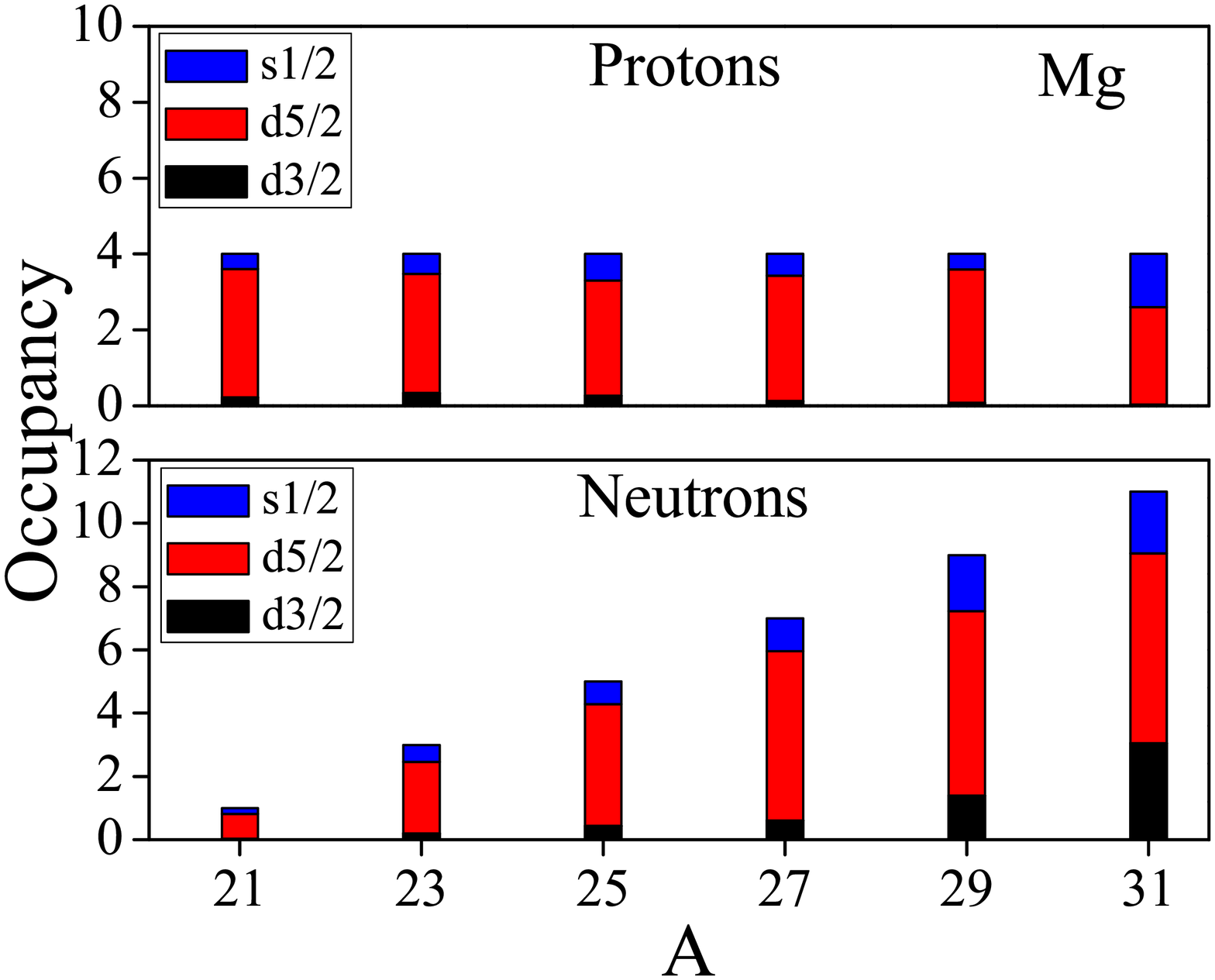}
\includegraphics[width=8cm,height=6.0cm,clip]{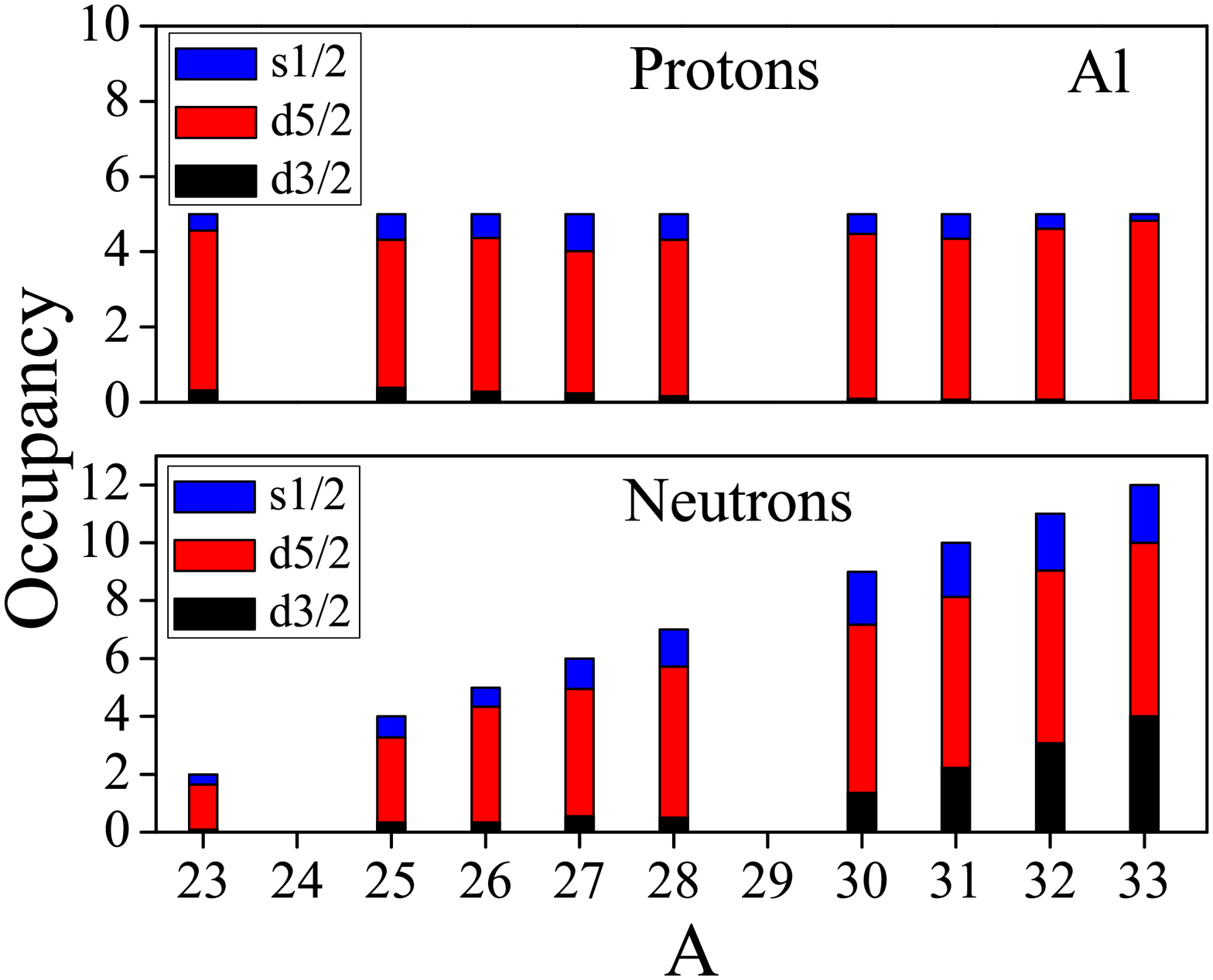}
\includegraphics[width=8cm,height=6cm,clip]{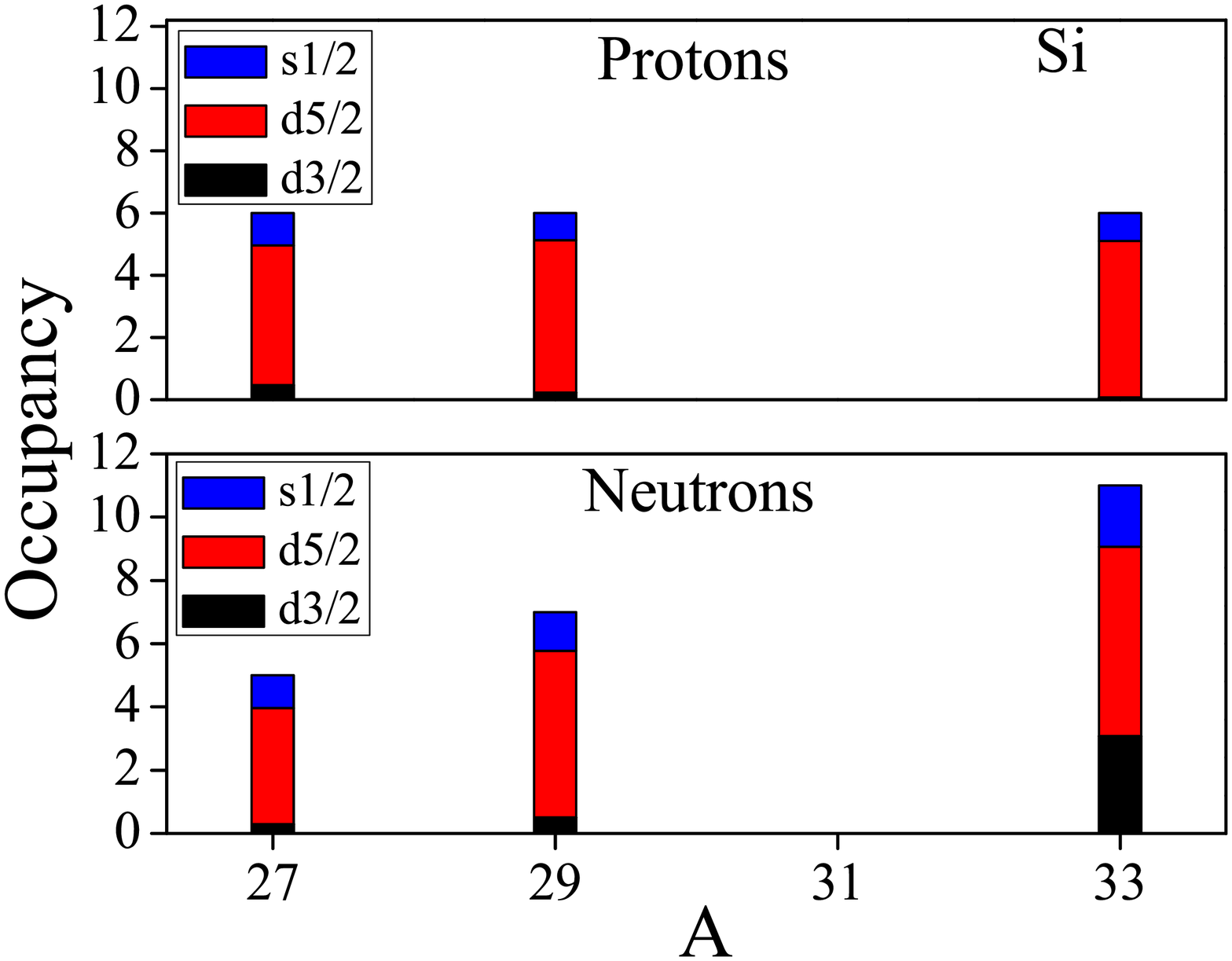}
\includegraphics[width=8cm,height=6cm,clip]{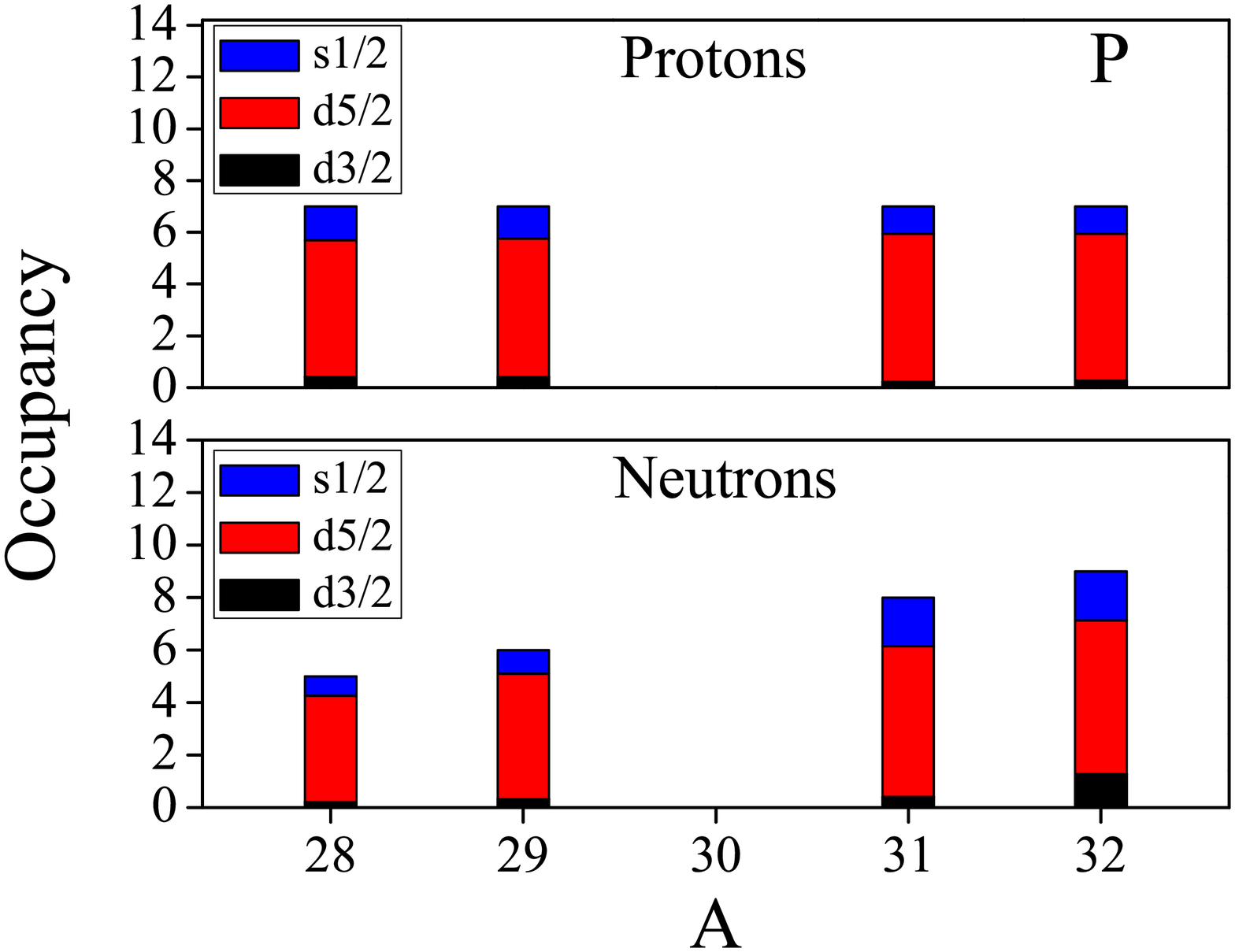}
\caption { {\label{Occu}Occupancies of $d_{3/2}$, $d_{5/2}$ and $s_{1/2}$ proton and neutron orbitals for $^{19-22}$F, $^{19-23}$Ne, $^{20-31}$Na, $^{21-31}$Mg, $^{23,25-28,30-33}$Al,  $^{27,29,33}$Si
and $^{28-29,31-32}$P  isotopes with CCEI for the g.s.
We have reported occupancies of those nuclei for which the quadrupole/magnetic moments are calculated in the present work.}}

\end{figure*}

For the oxygen isotopes, the calculated values of the
magnetic moments with IM-SRG and CCEI show almost
similar results. We have reported the magnetic moments
for $^{17-20}$O and quadrupole moments for $^{17-19}$O isotopes.
For $^{17}$O, the calculated magnetic moment and quadrupole
moment values using all the interactions are similar, because this is the single-particle moment (similarly for $^{17}$F).
The calculated magnetic moment of the $4_1^+$  state of $^{18}$O and of the  $5/2^+$ (g.s.) for $^{19}$O
and $2_1^+$  state of $^{20}$O are showing negative sign, while the sign has not yet been
confirmed experimentally. For $^{19}$O, both $ab~initio$ interactions give the opposite sign of the quadrupole moment with
USDB interaction; however, the sign from an experiment is
not yet confirmed. The calculated value of magnetic moments for $^{17,18,20}$F isotopes (g.s. and some isomers) are
close to the experimental data. However, $ab~initio$ 
interactions give slightly different values in comparison to the experimental
data for $^{19,21}$F. In the case of quadrupole moments, experimental sign of $^{17-22}$F isotopes are not yet confirmed. All
the interactions give the same sign.

The experimental data for magnetic moments of
Ne isotopes are available from $^{19}$Ne to $^{25}$Ne, while the
quadrupole moments are available for $^{21,22,23}$Ne. In Ref.
\cite{ne_geithner}, shell-model results of magnetic moments are reported
with USD and CW interactions for odd $^{23-25}$Ne isotopes.
Shape changes are reported to occur from collective to
single particle in Ne isotopes when moving from $^{19}$Ne to $^{25}$Ne. 
It is shown that the magnetic moment is more sensitive than the quadrupole moment for deciding the
structure of the nucleus. Our calculated results using
$ab~initio$ approaches for magnetic moments are showing
reasonable agreement with the experimental data except for
$^{23,25}$Ne isotopes. We have reported the quadrupole moments for $^{20-23}$Ne isotopes. Both $ab~initio$ results are in
good agreement with the experimental data. In the case
of $^{20}$Ne, the calculated $Q_{2^+}$ with all the three interactions
is approximately $-$0.15 eb, while experimental value is $-$0.23(3) eb.

In the $^{26-31}$Na chain \cite{na}, it is claimed that the experimental results of the magnetic moments for $^{26-29}$Na are
well described with $sd$ model space using the USD Hamiltonian.
The disagreement appears for the $^{30-31}$Na magnetic moments.
The $^{30,31}$Na isotopes are suggested to be
members of the island of inversion, as shown by Utsuno
$et al.$ \cite{otsuno01}. We have calculated the magnetic moments
for the g.s. for $^{20-31}$Na isotopes. In the case of $^{21,22}$Na and $^{24}$Na,
the magnetic moments are also given for the
first excited state. The results obtained from $ab~initio$
approaches for the g.s. are in reasonable agreement with
the experimental values (Fig. 1). In the case of $^{24}$Na,
for the magnetic moment (for the first excited state),
the sign is reverse using IM-SRG while CCEI gives the
sign which is in agreement with the experimental data.
We have calculated the quadrupole moments for $^{20-23}$Na
and $^{25-31}$Na. For $^{26}$Na, $ab~initio$ IM-SRG interaction is
giving same sign of the quadrupole moment as in the
phenomenological effective interaction. Experimentally,
the sign has not yet been confirmed. In the case of $^{30}$Na,
the results obtained from theory are far from the experimental value because $^{30}$Na is an element of the island
of inversion. To explain the quadrupole moment for $^{30}$Na
we need $pf$ model space. Utsuno $et al.$ \cite{otsuno01}, performed
theoretical calculations in $sdf_{7/2}p_{3/2}$ model space using
SDPF-M interaction \cite{SDPF-M} for $^{27,29}$Na isotopes using the
Monte Carlo shell-model approach.

For Mg isotopes, the $sd$  shell-model space is able to explain the experimental data reasonably well up to $^{29}$Mg
with all the three interactions. However, in the case of $^{27}$Mg, 
IM-SRG interaction gives opposite sign with the experimental data. In the case of $^{23}$Mg, for $Q(3/2^+)$, the 
experimental sign has not yet been confirmed. Recently, the sign of magnetic moment has been measured using laser spectroscopy at CERN-ISOLDE \cite{gerda_29Na}.
Shell model calculations predict a positive sign for the quadrupole moment.
In the case of $^{27}$Mg, the negative magnetic moment for
the ground state is dominated by the $\nu s_{1/2}$ configuration.
The IM-SRG fails to reproduce the correct sign of its magnetic moment, although it is also predicting $\nu s_{1/2}$ configuration for the ground state.
For $^{29}$Mg, the g.s. spin is
$I=3/2$ ($\nu d_{3/2}$), and all interactions give the correct sign
of the magnetic moment. The $sd$ model space is not able
to reproduce correctly the measured ground state $1/2^+$ for $^{31}$Mg. 
For this isotope, a strongly prolate deformed
ground state is reported in Ref. \cite{mg}. Recent theoretical
results reveal the existence of $2p-2h$ and $4p-4h$ configurations for $^{32}$Mg \cite{Macchiavelli}.

For $^{25-28,30-33}$Al isotopes, the $sd$ model space is able
to correctly reproduce the magnetic moments. In most
of the isotopes, CCEI results are in reasonable agreement
with the experimental data. For $^{25,28,30,32}$Al isotopes,
the sign for the g.s. magnetic moment is not yet confirmed, theoretically all the three interactions predict positive sign.
The calculated $\mu(2^+)$ and $\mu(3^+)$ for $^{28}$Al with IM-SRG
and CCEI are not showing good agreement with the experimental data. Also, $ab~initio$ interactions are giving
smaller values of the magnetic moment for $^{27}$Al. Experimentally, the sign of the quadrupole moment is confirmed only for
$^{26,27}$Al isotopes. For $^{26}$Al, the CCEI result for quadrupole
moment is in reasonable agreement with experimental
data, while for $^{27}$Al, the result of quadrupole moment
with IM-SRG is better than CCEI.

In the Si chain, the magnetic moments for $^{27,29,33}$Si isotopes are calculated for the g.s. while for $^{28,30}$Si isotopes
they are for the first excited state. For $^{28,29}$Si isotopes, the magnetic moments from the CCEI approach are in reasonable
agreement. The calculated $\mu(1/2^+)$ for $^{29}$Si is suppressed with IM-SRG in comparison with the experimental
data. The $ab~initio$ interactions are giving smaller values
with the opposite sign of $\mu(5/2^+)$ for $^{27}$Si in comparison to the
experimental data and with USDB interaction; however, the
experimental sign is tentative. The quadrupole moment
is calculated for $^{27}$Si, $^{28}$Si, and $^{30}$Si. For $^{30}$Si, all the
three interactions fail to reproduce the correct sign of
the quadrupole moment.

We have also reported the magnetic moments of
$^{28,29,31,32}$P, and all the calculated results support positive signs
for P isotopes in the g.s. except $^{32}$P. For $^{32}$P, IM-SRG
supports the experimental sign but the magnitude is larger
in comparison with the experimental value. We have also
calculated the magnetic moment of $^{31,32}$S and the quadrupole
moment of $^{32,33}$S. The CCEI results for magnetic moments ($^{32}$S) and quadrupole moments ($^{33}$S) are in reasonable agreement with the experimental data.

The calculated $g$ factors with $ab~ initio$ interactions for
the yrast levels in even-even $N = Z$ nuclei are $\sim$ 0.5.
Thus, the calculated value is in agreement with the experimental
value as reported in Ref. \cite{Kusoglu}.

In Fig. \ref{Mag}, we have shown the comparison between
the experimental and theoretical magnetic dipole moments for F, Ne, Na, Mg, Al, Si, and P isotopes. From this
figure, it is clear that $ab~initio$ interactions are not giving value close to the experimental data for heavier
$Z$ nuclei. The deviation between the calculated and the
experimental data is large for P isotopes. In the Figs. \ref{Mag} and \ref{Quad}, we have shown comparison only for the g.s. 
of those nuclei which have confirmed experimental signs. Apart from this, we have 
also plotted the g.s. of those data for which all the interactions are giving the same sign but their experimental signs are not yet confirmed.

The g.s. quadrupole moments for F, Ne, Na, Mg and Al isotopes are shown in the Fig. \ref{Quad}. All the interactions are
giving reasonable results for the F isotopes. For the
$^{22}$Na, the calculated quadrupole moment is larger in comparison with the experimental data for all the three interactions. For $^{25}$Na, $ab~initio$ interactions give larger
$Q(5/2^+)$ values in comparison with the experimental data;
however, the result with USDB interaction is reasonable.

The occupancies of $d_{3/2}$, $d_{5/2}$, and $s_{1/2}$ proton and neutron orbitals for $^{19-22}$F, $^{19-23}$Ne, $^{20-31}$Na, $^{21-31}$Mg, $^{23,25-28,30-33}$Al,  $^{27,29,33}$Si
and $^{28-29,31-32}$P  isotopes with CCEI for the g.s  are
shown in Fig. \ref{Occu}. In general, the role of the $d_{5/2}$ orbital is important as neutron number increases.

For $^{30}$Na (expt. g.s. is $2^+$), IM-SRG and CCEI effective interactions predict g.s. as $0^+$, while USDB predicts $2^+$.
For $^{30}$Na, the calculated magnetic moment with IM-SRG
interaction is in reasonable agreement with the experimental data; however, all three interactions give opposite signs for quadrupole moment in comparison with
the experimental data. For $^{31}$Na (experimental g.s. is $3/2^+$),
CCEI predicts the correct g.s. while USDB and IM-SRG give
$5/2^+$. For this nuclei, the magnetic moments predicted by
$ab~initio$ interactions are similar, whereas the quadrupole
moment predicted by CCEI is close to the experimental
data. For $^{31}$Mg (experimental g.s. is $1/2^+$ ), $ab~initio$ and USDB
interactions give g.s. of $3/2^+$. The calculated magnetic moment with CCEI is far from the experimental value and
also the sign is not correct. For $^{33}$Al, all the three interactions give the g.s. as $5/2^+$, although experimentally it is not
yet confirmed. The IM-SRG gives value of magnetic moment close to the experimental data for this isotope. In the
case of quadrupole moment, sign is not yet confirmed experimentally, but the magnitude is in reasonable agreement
with experimental value with all the three interactions.

The wave functions of the nuclei which show disagreement between $ab ~ initio$ results and with the experimental data and USDB results are shown in Table \ref{wave}.
 For $^{27}{\rm Al}$, both $ab~initio$ interactions give same structure. The $ab~initio$ results for $\mu_{5/2^+}$ are not in a good agreement with the experimental data.
In $^{28}{\rm Al}$, the CCEI result for $\mu_{2_1^+}$ is very far from IM-SRG and USDB interactions as well as with the experimental data; also, the
wave function is different from IM-SRG and USDB. In the case of CCEI interaction for the $^{28}{\rm Al}$($\mu_{2_1^+})$ result, we have one unpaired proton and one unpaired neutron 
in $s_{1/2}$ orbits whereas in USDB and IM-SRG interactions, we have
one unpaired  neutron in $s_{1/2}$ orbits, and one unpaired proton in $d_{5/2}$ orbit. Although, USDB and IM-SRG are also not in very good agreement with the experimental data.
In $^{32}{\rm Al}$, all the three interactions
give same structure for the wave function but still CCEI result for $Q_{1_1^+}$ is far from the experimental data. 
In $^{27}{\rm Si}$, the structures of wave function for CCEI interaction is due to
two unpaired protons and one unpaired neutron, whereas in IM-SRG and USDB interactions come from one unpaired neutron in $d_{5/2}$ orbit.
In the case of $^{30}{\rm Si}$, the CCEI result for $Q_{2_1^+}$ is very far from the experiment, the structure coes because of the two unpaired protons which are in
$s_{1/2}$ and $d_{5/2}$ orbits, while for IM-SRG and USDB interactions two unpaired neutrons are in $d_{3/2}$ and $s_{1/2}$ orbits.
For $^{31}{\rm P}$ ($\mu_{1/2_1^+}$, $\mu_{5/2_1^+}$), $^{32}{\rm S}$ ($Q_{2_1^+}$) and $^{33}{\rm S}$ ($Q_{3/2_1^+}$),
the structures of wave functions for IM-SRG are very different from USDB and CCEI interactions. 
For these nuclei, we can see results from Tables \ref{mag} and \ref{qad} that show
the IM-SRG results are very far from the experimental data. For $^{28}{\rm Na}$, $^{25}{\rm Mg}$, $^{32}{\rm Al}$, $^{29}{\rm Si}$, and $^{29}{\rm P}$ all
three calculations give the same structure of the wave functions. But we can see from Table \ref{mag} a deviation of one of the $ab~initio$ results with the other two
interactions and the experimental data. For $^{28}{\rm P}$, all
the three calculations give same structure but the magnitude of USDB result is closer to the experimental data. However, the experimental sign has not yet been confirmed.

\begin{sidewaystable*}
\vspace{6cm}
  \begin{center}
    \caption{{\label{wave} Dominant configuration of the wave functions with  $ab~intio$ effective interactions and USDB effective interaction. 
    In these nuclei $ab~ initio$ results are
    showing deviation with experimental data and USDB effective interaction results.}} 
   \begin{tabular}{ l l l l l l l l }

\hline
\hline
Nuclei  &  Spin  & \hspace{10mm}IM-SRG     &  Probability & \hspace{12mm} CCEI         & Probability & \hspace{12mm}   USDB       & Probability   \\
\hline

$^{28}{\rm Na}$  & $\mu_{1^+}$    &$\pi (d_{3/2}^0$,$d_{5/2}^3$,$s_{1/2}^0$) $\otimes$ $\nu (d_{3/2}^1$,$d_{5/2}^6$,$s_{1/2}^2)$ & 41.71\%  &  $\pi(d_{3/2}^0$,$d_{5/2}^3$,$s_{1/2}^0$) $\otimes$ $\nu(d_{3/2}^1$,$d_{5/2}^6$,$s_{1/2}^2$)  & 43.52\% &$\pi (d_{3/2}^0$,$d_{5/2}^3$,$s_{1/2}^0$) $\otimes$ $\nu (d_{3/2}^1$,$d_{5/2}^6$,$s_{1/2}^2)$   &  51.85\%    \\
$^{25}{\rm Mg}$  & $\mu_{5/2_1^+}$    &$\pi (d_{3/2}^0$,$d_{5/2}^4$,$s_{1/2}^0$) $\otimes$ $\nu (d_{3/2}^0$,$d_{5/2}^5$,$s_{1/2}^0)$ & 16.99\% &  $\pi(d_{3/2}^0$,$d_{5/2}^4$,$s_{1/2}^0$) $\otimes$ $\nu(d_{3/2}^0$,$d_{5/2}^5$,$s_{1/2}^0$)  & 18.14\%  &$\pi (d_{3/2}^0$,$d_{5/2}^4$,$s_{1/2}^0$) $\otimes$ $\nu (d_{3/2}^0$,$d_{5/2}^5$,$s_{1/2}^0)$   &  26.20\%    \\
$^{27}{\rm Al}$  & $\mu_{5/2_1^+}$   &$\pi (d_{3/2}^0$,$d_{5/2}^4$,$s_{1/2}^1$) $\otimes$ $\nu (d_{3/2}^0$,$d_{5/2}^5$,$s_{1/2}^1)$ & 6.46\%  &  $\pi(d_{3/2}^0$,$d_{5/2}^4$,$s_{1/2}^1$) $\otimes$ $\nu(d_{3/2}^0$,$d_{5/2}^5$,$s_{1/2}^1$)  & 12.92\%  &$\pi (d_{3/2}^0$,$d_{5/2}^5$,$s_{1/2}^0$) $\otimes$ $\nu (d_{3/2}^0$,$d_{5/2}^6$,$s_{1/2}^0)$   &  27.19\%    \\
$^{28}{\rm Al}$  & $\mu_{3_1^+}$   &$\pi (d_{3/2}^0$,$d_{5/2}^5$,$s_{1/2}^0$) $\otimes$ $\nu (d_{3/2}^0$,$d_{5/2}^6$,$s_{1/2}^1)$ & 14.98\%  &  $\pi(d_{3/2}^0$,$d_{5/2}^5$,$s_{1/2}^0$) $\otimes$ $\nu(d_{3/2}^0$,$d_{5/2}^6$,$s_{1/2}^1$)  & 29.47\%&$\pi (d_{3/2}^0$,$d_{5/2}^5$,$s_{1/2}^0$) $\otimes$ $\nu (d_{3/2}^0$,$d_{5/2}^6$,$s_{1/2}^1)$   &  37.28\%     \\
                 & $\mu_{2_1^+}$  &$\pi (d_{3/2}^0$,$d_{5/2}^5$,$s_{1/2}^0$) $\otimes$ $\nu (d_{3/2}^0$,$d_{5/2}^6$,$s_{1/2}^1)$ & 7.72\% &  $\pi(d_{3/2}^0$,$d_{5/2}^4$,$s_{1/2}^1$) $\otimes$ $\nu(d_{3/2}^0$,$d_{5/2}^6$,$s_{1/2}^1$)  & 11.66\%  &$\pi (d_{3/2}^0$,$d_{5/2}^5$,$s_{1/2}^0$) $\otimes$ $\nu (d_{3/2}^0$,$d_{5/2}^6$,$s_{1/2}^1)$   &  28.15\%     \\                
$^{32}{\rm Al}$  & $Q_{1_1^+}$   &$\pi (d_{3/2}^0$,$d_{5/2}^5$,$s_{1/2}^0$) $\otimes$ $\nu (d_{3/2}^3$,$d_{5/2}^6$,$s_{1/2}^2)$ & 71.50\% &  $\pi(d_{3/2}^0$,$d_{5/2}^5$,$s_{1/2}^0$) $\otimes$ $\nu(d_{3/2}^3$,$d_{5/2}^6$,$s_{1/2}^2$)  & 65.89\%   &$\pi (d_{3/2}^0$,$d_{5/2}^5$,$s_{1/2}^0$) $\otimes$ $\nu (d_{3/2}^3$,$d_{5/2}^6$,$s_{1/2}^2)$   &  78.19\%    \\
                                
$^{27}{\rm Si}$  & $\mu_{5/2_1^+}$   &$\pi (d_{3/2}^0$,$d_{5/2}^6$,$s_{1/2}^0$) $\otimes$ $\nu (d_{3/2}^0$,$d_{5/2}^5$,$s_{1/2}^0)$ & 9.27\%  &  $\pi(d_{3/2}^0$,$d_{5/2}^5$,$s_{1/2}^1$) $\otimes$ $\nu(d_{3/2}^0$,$d_{5/2}^4$,$s_{1/2}^1$)  & 13.24\%   &$\pi (d_{3/2}^0$,$d_{5/2}^6$,$s_{1/2}^0$) $\otimes$ $\nu (d_{3/2}^0$,$d_{5/2}^5$,$s_{1/2}^0)$   &  27.35\%    \\                 
$^{29}{\rm Si}$  & $\mu_{1/2_1^+}$  &$\pi (d_{3/2}^0$,$d_{5/2}^6$,$s_{1/2}^0$) $\otimes$ $\nu (d_{3/2}^0$,$d_{5/2}^6$,$s_{1/2}^1)$ & 11.62\%   &  $\pi(d_{3/2}^0$,$d_{5/2}^6$,$s_{1/2}^0$) $\otimes$ $\nu(d_{3/2}^0$,$d_{5/2}^6$,$s_{1/2}^1$)  & 29.18\%  &$\pi (d_{3/2}^0$,$d_{5/2}^6$,$s_{1/2}^0$) $\otimes$ $\nu (d_{3/2}^0$,$d_{5/2}^6$,$s_{1/2}^1)$   &  32.52\%    \\ 
  
 $^{30}{\rm Si}$  & $Q_{2_1^+}$  &$\pi (d_{3/2}^0$,$d_{5/2}^6$,$s_{1/2}^0$) $\otimes$ $\nu (d_{3/2}^1$,$d_{5/2}^6$,$s_{1/2}^1)$ & 8.21\%   &  $\pi(d_{3/2}^0$,$d_{5/2}^5$,$s_{1/2}^1$) $\otimes$ $\nu(d_{3/2}^0$,$d_{5/2}^6$,$s_{1/2}^2$)  & 33.80\%  &$\pi (d_{3/2}^0$,$d_{5/2}^6$,$s_{1/2}^0$) $\otimes$ $\nu (d_{3/2}^1$,$d_{5/2}^6$,$s_{1/2}^1)$   &  25.39\%    \\ 
 
$^{28}{\rm P}$  & $\mu_{3_1^+}$  &$\pi (d_{3/2}^0$,$d_{5/2}^6$,$s_{1/2}^1$) $\otimes$ $\nu (d_{3/2}^0$,$d_{5/2}^5$,$s_{1/2}^0)$ & 8.61\%  &  $\pi(d_{3/2}^0$,$d_{5/2}^6$,$s_{1/2}^1$) $\otimes$ $\nu(d_{3/2}^0$,$d_{5/2}^5$,$s_{1/2}^0$)  & 29.43\%  &$\pi (d_{3/2}^0$,$d_{5/2}^6$,$s_{1/2}^1$) $\otimes$ $\nu (d_{3/2}^0$,$d_{5/2}^5$,$s_{1/2}^0)$   &  37.24\%    \\
  
$^{29}{\rm P}$  & $\mu_{1/2_1^+}$   &$\pi (d_{3/2}^0$,$d_{5/2}^6$,$s_{1/2}^1$) $\otimes$ $\nu (d_{3/2}^0$,$d_{5/2}^6$,$s_{1/2}^0)$ & 10.38\%  &  $\pi(d_{3/2}^0$,$d_{5/2}^6$,$s_{1/2}^1$) $\otimes$ $\nu(d_{3/2}^0$,$d_{5/2}^6$,$s_{1/2}^0$)  & 28.56\%  &$\pi (d_{3/2}^0$,$d_{5/2}^6$,$s_{1/2}^1$) $\otimes$ $\nu (d_{3/2}^0$,$d_{5/2}^6$,$s_{1/2}^0)$   &  32.60\%    \\ 

$^{31}{\rm P}$  & $\mu_{1/2_1^+}$    &$\pi (d_{3/2}^0$,$d_{5/2}^6$,$s_{1/2}^1$) $\otimes$ $\nu (d_{3/2}^2$,$d_{5/2}^6$,$s_{1/2}^0)$ & 6.46\% &  $\pi(d_{3/2}^0$,$d_{5/2}^6$,$s_{1/2}^1$) $\otimes$ $\nu(d_{3/2}^0$,$d_{5/2}^6$,$s_{1/2}^2$)  & 66.90\%  &$\pi (d_{3/2}^0$,$d_{5/2}^6$,$s_{1/2}^1$) $\otimes$ $\nu (d_{3/2}^0$,$d_{5/2}^6$,$s_{1/2}^2)$   & 34.38\%    \\
                & $\mu_{3/2_1^+}$   &$\pi (d_{3/2}^0$,$d_{5/2}^6$,$s_{1/2}^1$) $\otimes$ $\nu (d_{3/2}^2$,$d_{5/2}^6$,$s_{1/2}^0)$ & 6.44\%  &  $\pi(d_{3/2}^0$,$d_{5/2}^6$,$s_{1/2}^1$) $\otimes$ $\nu(d_{3/2}^1$,$d_{5/2}^6$,$s_{1/2}^1$)  & 22.13\%  &$\pi (d_{3/2}^1$,$d_{5/2}^6$,$s_{1/2}^0$) $\otimes$ $\nu (d_{3/2}^0$,$d_{5/2}^6$,$s_{1/2}^2)$   &  15.28\%    \\
                & $\mu_{5/2_1^+}$  &$\pi (d_{3/2}^1$,$d_{5/2}^5$,$s_{1/2}^1$) $\otimes$ $\nu (d_{3/2}^2$,$d_{5/2}^6$,$s_{1/2}^0)$ & 5.94\%   &  $\pi(d_{3/2}^0$,$d_{5/2}^5$,$s_{1/2}^2$) $\otimes$ $\nu(d_{3/2}^0$,$d_{5/2}^6$,$s_{1/2}^2$)  & 53.89\%  &$\pi (d_{3/2}^0$,$d_{5/2}^5$,$s_{1/2}^2$) $\otimes$ $\nu (d_{3/2}^0$,$d_{5/2}^6$,$s_{1/2}^2)$   &  19.05\%    \\
                
$^{32}{\rm S}$  & $Q_{2_1^+}$    &$\pi (d_{3/2}^2$,$d_{5/2}^5$,$s_{1/2}^1$) $\otimes$ $\nu (d_{3/2}^2$,$d_{5/2}^5$,$s_{1/2}^1)$ & 4.72\%  &  $\pi(d_{3/2}^0$,$d_{5/2}^6$,$s_{1/2}^2$) $\otimes$ $\nu(d_{3/2}^1$,$d_{5/2}^6$,$s_{1/2}^1$)  & 36.14\% &$\pi (d_{3/2}^0$,$d_{5/2}^6$,$s_{1/2}^2$) $\otimes$ $\nu (d_{3/2}^1$,$d_{5/2}^6$,$s_{1/2}^1)$   &  11.36\%    \\                
                
$^{33}{\rm S}$  & $Q_{3/2_1^+}$  &$\pi (d_{3/2}^2$,$d_{5/2}^6$,$s_{1/2}^0$) $\otimes$ $\nu (d_{3/2}^1$,$d_{5/2}^6$,$s_{1/2}^2)$ & 7.93\%   &  $\pi(d_{3/2}^0$,$d_{5/2}^6$,$s_{1/2}^2$) $\otimes$ $\nu(d_{3/2}^1$,$d_{5/2}^6$,$s_{1/2}^2$)  & 77.18\%  &$\pi (d_{3/2}^0$,$d_{5/2}^6$,$s_{1/2}^2$) $\otimes$ $\nu (d_{3/2}^1$,$d_{5/2}^6$,$s_{1/2}^2)$   & 47.69\%    \\                
                
\hline\hline
\end{tabular}
\end{center}
\label{partition}
\end{sidewaystable*}
\hspace{-1mm}
\section{Summary}

In the present work using two $ab ~ initio$ approaches we have reported the quadrupole and
magnetic moments  for $sd$ shell nuclei with the shell model.
We perform calculations with effective interactions derived from in-medium similarity renormalization and coupled-cluster approaches.
Along with $ab ~ initio$ interactions, we have also compared these results with the phenomenological USDB interaction.
The results show reasonable agreement with the available experimental data. This work will add more information to
previously known spectroscopic properties of $sd$ shell nuclei from $ab ~ initio$ approaches.

\section*{Acknowledgment:}
We would like to thank  Gerda Neyens  for many useful suggestions and comments on this manuscript.
P.C.S. would like to thank S. R. Stroberg for discussions on IM-SRG.
A.S. acknowledges financial support from MHRD ( Government of India) for her Ph.D. thesis work.
P.C.S. acknowledges financial support from faculty initiation grants.



\begin{thebibliography}{37}
\expandafter\ifx\csname natexlab\endcsname\relax\def\natexlab#1{#1}\fi
\expandafter\ifx\csname bibnamefont\endcsname\relax
  \def\bibnamefont#1{#1}\fi
\expandafter\ifx\csname bibfnamefont\endcsname\relax
  \def\bibfnamefont#1{#1}\fi
\expandafter\ifx\csname citenamefont\endcsname\relax
  \def\citenamefont#1{#1}\fi
\expandafter\ifx\csname url\endcsname\relax
  \def\url#1{\texttt{#1}}\fi
\expandafter\ifx\csname urlprefix\endcsname\relax\def\urlprefix{URL }\fi
\providecommand{\bibinfo}[2]{#2}
\providecommand{\eprint}[2][]{\url{#2}}


\bibitem{otsu1}
T.~ Otsuka, T. Suzuki, J.D. Holt, A. Schwenk, and Y. Akaisji, Three-Body Forces and the Limit of Oxygen Isotopes,
{\color {blue} Phys. Rev. Lett. \textbf{105}, 032501 (2010)}.


\bibitem{otsu2}
J. D. Holt, T. Otsuka, A. Schwenk, and T. Suzuki, Three-body forces and shell structure in calcium isotopes, {\color {blue} J. Phys. G \textbf{39}, 085111 (2012)}.

\bibitem{stroberg}
S. R. Stroberg, H. Hergert, J. D. Holt, S. K. Bogner, and A. Schwenk,  Ground and excited states of doubly open-shell nuclei from ab initio valence-space Hamiltonians,
{\color {blue} Phys. Rev. C \textbf{93}, 051301(R) (2016)}.

 \bibitem{jan1}
G. R. Jansen, M. D. Schuster, A. Signoracci, G. Hagen, and P. Navr\'atil, Open $sd$-shell nuclei from first principles,  {\color {blue} Phys. Rev. C \textbf{94}, 011301(R) (2016)}.


\bibitem{jan2}
G. R. Jansen, J. Engel, G. Hagen, P. Navratil, and A. Signoracci,
$Ab~Initio$ Coupled-Cluster Effective Interactions for the Shell Model: Application to Neutron-Rich Oxygen and Carbon Isotopes,
{\color {blue} Phys. Rev. Lett. \textbf{113},  142502  (2014)}.

 \bibitem{sri}
 P.C. Srivastava and V. Kumar, Spectroscopic factor strengths using $ab~initio$ approaches, {\color {blue} Phys. Rev. C \textbf{94}, 064306 (2016)}.


 \bibitem{Dikmen}
E. Dikmen, A. F. Lisetskiy, B. R. Barrett, P. Maris, A. M. Shirokov, and J. P. Vary,  $Ab~initio$ effective interactions for $sd$-shell valence nucleons,
{\color {blue} Phys. Rev. C \textbf{91}, 064301 (2015)}.


\bibitem{24F_imsrg}
L. Cáceres, A. Lepailleur, O. Sorlin, M. Stanoiu, D. Sohler, Z.
Dombradi, S. K. Bogner, B. A. Brown, H. Hergert, J. D. Holt
{\it et al.}, Nuclear structure studies of $^{24}$F, {\color {blue} Phys. Rev. C \textbf{92}, 014327 (2015)}.

\bibitem{25F_cc}
Z. Vajta, M. Stanoiu, D. Sohler, G. R. Jansen, F. Azaiez, Z.
Dombradi, O. Sorlin, B. A. Brown, M. Belleguic, C. Borcea
{\it et al.}, Excited states in the neutron-rich nucleus $^{25}\mathrm{F}$, {\color {blue} Phys. Rev. C \textbf{89}, 054323 (2014)}.

 
\bibitem{usdb}
B. A. Brown and W. A. Richter, New ``USD'' Hamiltonians for the $\mathit{sd}$ shell, {\color {blue} Phys. Rev. C \textbf{74}, 034315 (2006)}.


\bibitem{Nushellx} B. A. Brown and  W. D. M.  Rae, The Shell-Model Code NuShellX@MSU, {\color {blue} Nuclear Data Sheets {\bf 120}, 115 (2014)}.

\bibitem{bogner} S.K. Bogner, R.J. Furnstahl and A. Schwenk, From low-momentum interactions to nuclear structure, {\color {blue} Prog. Part. Nucl. Phys. \textbf{65},
  94 (2010)}.
  
 \bibitem{tsukiyama} K. Tsukiyama, S.K. Bogner and A. Schwenk, In-Medium Similarity Renormalization Group For Nuclei,
 {\color {blue} Phys. Rev. Lett. \textbf{106},  222502  (2011)}.

\bibitem{PRC78_064302}  W. A. Richter, S. Mkhize, and B. Alex Brown, $sd$-shell observables for the USDA and USDB Hamiltonians,
{\color {blue} Phys. Rev. C \textbf{78}, 064302 (2008)}.

\bibitem{stone2005} N. J. Stone,  {\color {blue} Atomic Data and Nuclear Data Tables  {\bf 90}, 75 (2005)}.

\bibitem{IAEA https://www-nds.iaea.org/nuclearmoments/} {\color {blue} IAEA https://www-nds.iaea.org/nuclearmoments/}.

\bibitem{DeRydt2013} M. De Rydt, M. Depuydt and G. Neyens, {\color {blue} Atomic Data and Nuclear Data Tables  {\bf 99}, 391 (2013)}.

\bibitem{ne} 
H. Iwasaki, T. Motobayashi, H. Sakurai, K. Yoneda, T. Gomi, N. Aoi, N. Fukuda, Zs. Fülöp, U. Futakami, Z. Gacsi {\it et al.}, Quadrupole collectivity of $^{28}$Ne 
and the boundary of the island of inversion,
{\color {blue} Phys. Lett. B \textbf{620}, 118 (2005)}.

\bibitem{ne1} 
 B.V. Pritychenko, T. Glasmacher, P.D. Cottle, M. Fauerbach, R.W. Ibbotson, K.W. Kemper, V. Maddalena, A. Navin, R. Ronningen, A. Sakharuk {\it et al.},
 Role of intruder configurations in $^{26,28}$Ne and $^{30,32}$Mg,
 {\color {blue} Phys. Lett. B \textbf{461}, 322 (1999)}.

\bibitem{mg}
G. Neyens, M. Kowalska, D. Yordanov, K. Blaum, P. Himpe, P.
Lievens, S. Mallion, R. Neugart, N. Vermeulen, Y. Utsuno, and
T. Otsuka, {\it et al.}, Measurement of the Spin and Magnetic Moment of $^{31}\mathrm{M}\mathrm{g}$: Evidence for a Strongly Deformed Intruder Ground State,
{\color {blue} Phys. Rev. Lett. \textbf{94},
  022501 (2005)}.

 
  

\bibitem{al}
M. De Rydt, G. Neyens, K. Asahi, D.L. Balabanski, J.M. Daugas, M. Depuydt, L. Gaudefroy, S. Gr\'evy, Y. Hasama, Y. Ichikawa {\it et al.},
Precision measurement of the electric quadrupole moment of $^{31}$Al and determination of the effective proton charge in the sd-shell,
{\color {blue} Phys. Lett. B \textbf{678}, 344 (2009)}.


\bibitem{al1} 
P. Himpe, G. Neyens, D.L. Balabanski, G. Bélier, D. Borremans, J. M. Daugas, F. de Oliveira Santos, M. De Rydt,
K. Flanagan, G. Georgiev {\it et al.}, g factors of $^{31,32,33}$Al: Indication for intruder configurations in the $^{33}$Al ground state,
{\color {blue} Phys. Lett. B \textbf{643}, 257 (2006)}.


\bibitem{na} 
M. Keim, U. Georg, A. Klein, R. Neugart, M. Neuroth, S. Wilbert, P. Lievens, L. Vermeeren, B. A. Brown, and ISOLDE Collaboration, 
Measurement of the electric quadrupole moments of $^{26–29}$Na,
{\color {blue} Eur. Phys. J. A \textbf{8}, 31 (2000)}.
  
\bibitem{30ne_be2} 
P. Doornenbal, H. Scheit, S. Takeuchi, N. Aoi, K. Li, M.
Matsushita, D. Steppenbeck, H. Wang, H. Baba, E. Ideguchi {\it et al.}, Mapping the deformation in the “island of inversion”: Inelastic scattering of $^{30}$Ne and 
$^{36}$Mg at intermediate energies,
{\color {blue} Phys. Rev. C \textbf{93}, 044306 (2016)}.


\bibitem{29na_be2} 
A.M. Hurst, C.Y. Wu, J.A. Becker, M.A. Stoyer, C.J. Pearson, G. Hackman, M.A. Schumaker, C.E. Svensson,
R.A.E. Austin, G.C. Ball {\it et al.}, Narrowing of the neutron $sd$-$pf$
shell gap in $^{29}$Na, {\color {blue} Phys. Lett. B \textbf{674}, 168 (2009)}.


\bibitem{2930na_be2} 
M. Seidlitz, P. Reiter, R. Altenkirch, B. Bastin, C. Bauer, A.
Blazhev, N. Bree, B. Bruyneel, P. A. Butler, J. Cederkall {\it et al.}, Coulomb excitation of $^{29,30}$Na: Mapping the borders of the island of inversion,
{\color {blue} Phys. Rev. C \textbf{89}, 024309 (2014)}.

\bibitem{30mg_be2} 
O. Niedermaier, H. Scheit, V. Bildstein, H. Boie, J. Fitting, R.von Hahn, F. Kock, M. Lauer, U.K. Pal,
H. Podlech {\it et al.}, ``Safe'' Coulomb Excitation of $^{30}$Mg, {\color {blue} Phys. Rev. Lett. \textbf{94}, 172501 (2005)}.



\bibitem{31mb_coulex} 
M. Seidlitz, D. Mücher, P. Reiter, V. Bildstein, A. Blazhev, N. Bree, B. Bruyneel, J. Cederkäll, 
E. Clement, T. Davinson {\it et al.}, Coulomb excitation of $^{31}$Mg, {\color {blue} Phys. Lett. B \textbf{700}, 181 (2011)}.



\bibitem{32mb_coulex} 
J. A. Church, C. M. Campbell, D.-C. Dinca, J. Enders, A. Gade,
T. Glasmacher, Z. Hu, R. V. F. Janssens, W. F. Mueller, H. Olliver {\it et al.},
Measurement of E2 transition strengths in $^{32,34}$Mg,
{\color {blue} Phys. Rev. C \textbf{72}, 054320 (2005)}.


\bibitem{31al_g} 
D. Borremans, S. Teughels, N.A. Smirnova, D.L. Balabanski, N. Coulier, J.-M Daugas, F. de Oliveira Santos, G. Georgiev, M. Lewitowicz, 
I. Matea {\it et al.}, Spin and magnetic moment of $^{31}$Al ground state, 
{\color {blue} Phys. Lett. B \textbf{537}, 45 (2002)}.


\bibitem{32al_q} 
D. Kameda, H. Ueno, K. Asahi, M. Takemura, A. Yoshimi, T. Haseyama, M. Uchida, K. Shimada, D. Nagae, G. Kijima  {\it et al.},
Measurement of the electric quadrupole moment of $^{32}$Al, {\color {blue} Phys. Lett. B \textbf{647}, 93 (2007)}.

\bibitem{33al_q}   
H. Heylen, M. De Rydt, G. Neyens, M. L. Bissell, L. Caceres, R.
Chevrier, J. M. Daugas, Y. Ichikawa, Y. Ishibashi, O. Kamalou {\it et al.}, 
High-precision quadrupole moment reveals significant intruder component in $^{33}_{13}$Al$_{20}$ ground state,
{\color {blue} Phys. Rev. C \textbf{94}, 034312 (2016)}.

\bibitem{PRC90_014302} 
E. Caurier, F. Nowacki and A. Poves, Merging of the islands of inversion at $N=20$ and $N=28$, 
{\color {blue} Phys. Rev. C \textbf{90}, 014302 (2014)}.

\bibitem{PRC70_044307} 
Y. Utsuno, T. Otsuka, T. Glasmacher, T. Mizusaki, and M. Honma, Onset of intruder ground state in exotic Na isotopes and evolution of the $N=20$ shell gap,
{\color {blue} Phys. Rev. C \textbf{70}, 044307 (2004)}.

\bibitem{be2} Data extracted using the NNDC World Wide Web site from the ENSDF database.

\bibitem{ne_geithner}
W. Geithner, B. A. Brown, K. M. Hilligsøe, S. Kappertz, M.
Keim, G. Kotrotsios, P. Lievens, K. Marinova, R. Neugart, H.
Simon, and S. Wilbert, Nuclear moments of neon isotopes in the range from $^{17}$Ne at the proton drip line to neutron-rich $^{25}$Ne, 
{\color {blue} Phys. Rev. C \textbf{71}, 064319  (2005)}.

\bibitem
{otsuno01} Y. Utsuno  {\it et al.}, Extreme location of F drip line and disappearance of the $N=20$ magic structure,
{\color {blue} Phys. Rev. C \textbf{64}, 011301(R) (2001)}.

\bibitem{Ne21}
D. Sundholm and J. Olsen, Finite element multiconfiguration Hartree-Fock calculations on carbon, oxygen, 
and neon: the nuclear quadrupole moments of carbon-11, oxygen-17, and neon-21, {\color {blue} J. Phys. Chem. \textbf{96}, 627 (1992)}.

\bibitem{stone1} N. J. Stone, {\color {blue} Atomic Data and Nuclear Data Tables  {\bf 111-112}, 1 (2016)}.

\bibitem{F19} A. Halkier, O. Christiansen, D. Sundholm, and P. Pyykkö, An improved value of the nuclear quadrupole moment of the 197 keV $I = 52$ excited state of $^{19}$F,
{\color {blue} Chem. Phys. Lett. \textbf{271}, 273 (1997)}.


\bibitem{Na23} 
P. Pyykko, and A. J. Sadlej, 
Determination of the 23Na nuclear quadrupole moment from molecular data for NaF and NaCl,
{\color {blue} Phys. Lett. \textbf{227}, 221 (1994)}.


\bibitem{epja}
G. Neyens, P. Himpe, D. L. Balabanski, P. Morel, L. Perrot, M. De Rydt, I. Stefan, C. Stodel, J. C. Thomas, N. Vermeulen {\it et al.}, 
The ``island of inversion'' from a nuclear moments perspective and the $g$ factor of $^{35}$Si,
{\color {blue} Eur. Phys. J. Special Topics {\bf 150}, 149 (2007)}.


\bibitem{Mg25} D. Sundholm and J. Olsen, Finite element MCHF calculations on Mg(3s3p; $^{3}$P$^{0}$):
The nuclear quadrupole moment of $^{25}$Mg, {\color {blue} Nucl. Phys. A \textbf{534}, 360 (1991)}.


\bibitem{Al27} 
V. Kell{\"o}, A. J. Sadlej, P. Pyykko, D. Sundholm, and  M. Tokman, 
Electric quadrupole moment of the $^{27}$Al nucleus: Converging results from the AlF and AlCl molecules and the Al atom, 
{\color {blue} Chem. Phys. Lett. \textbf{304}, 414 (1999)}.



\bibitem{Si27} 
K. Matsuta, T. Minamisono, M. Fukuda, M. Mihara, K. Sato,
K. Minamisono, T. Yamaguchi, T. Onishi, T. Miyake, M. Sasaki {\it et al.}, Recent studies on the nuclear moments of light mirror nuclei ($T=1/2, 3/2$),
{\color {blue} Nucl. Phys. A \textbf{704}, 98c (2002)}.
 


\bibitem{S33}  D. Sundholm and J. Olsen, Nuclear quadrupole moments of $^{33}$S and $^{35}$S, {\color {blue} Phys. Rev. A \textbf{42}, 1160 (1990)}.



\bibitem{SDPF-M}
Y. Utsuno, T. Otsuka, T. Mizusaki, and M. Honma, {\it et al.}, Varying shell gap and deformation in $N\sim20$ unstable nuclei studied by the Monte Carlo shell model,
 {\color {blue} Phys. Rev. C \textbf{60}, 054315 (1999)}.



\bibitem{gerda_29Na}
D. T. Yordanov, M. L. Bissell, K. Blaum, M. De Rydt, C.
Geppert, J. Kramer, K. Kreim, M. Kowalska, A. Krieger, P.
Lievens {\it et al.}, Spin and magnetic moment of $^{23}$Mg, {\color {blue} J. Phys. G \textbf{44}, 075104 (2017)}.



\bibitem{Macchiavelli} 
 A. O. Macchiavelli, H. L. Crawford, C. M. Campbell, R. M.
Clark, M. Cromaz, P. Fallon, M. D. Jones, I. Y. Lee, M. Salathe,
B. A. Brown, and A. Poves, The $^{30}$Mg($t,p$)$^{32}$Mg ``puzzle'' reexamined, {\color {blue} Phys. Rev. C \textbf{94},
  051303(R) (2016)}.
 
 

\bibitem{Kusoglu} 
A. Kusoglu, A. E. Stuchbery, G. Georgiev, B. A. Brown, A.
Goasduff, L. Atanasova, D. L. Balabanski, M. Bostan, M.
Danchev, P. Detistov {\it et al.}, Magnetism of an Excited Self-Conjugate Nucleus: Precise Measurement of the $g$ Factor of the $2_1^{+}$ State in $^{24}$Mg, {\color {blue}
Phys. Rev. Lett. \textbf{114}, 062501 (2015)}.



\end{thebibliography}
\end{document}